# A new migration mechanism for oversized solutes in cubic lattices: correlation effects

J.L. Bocquet [a], C. Barouh [b], Chu Chun Fu [b]
[a] CMLA, ENS Cachan, CNRS, Université Paris-Saclay, 94235 Cachan, France
[b] SRMP DEN Saclay, 91191 Gif-sur-Yvette, Cedex
jean-louis.bocquet@cmla.ens-cachan.fr

First-principles calculations confirmed recently that oversized solute atoms (OSA) are dissolved as substitutional species in BCC and FCC iron. The dilatational strain which accompanies their insertion on a lattice site attracts the vacancy defect (V). The new point is the strength of this attraction at first neighbour distance: for some metals, it is so large that the 1NN-pair OSA+V is no longer stable and relaxes spontaneously towards a new configuration where the OSA sits in the middle of the bond, the two ends of which are decorated with two half-vacancies (V/2) as depicted in Fig.1a-2a [1-3]. The complex is denoted (V/2+OSA+V/2) to remember that only one unoccupied vacancy was present before the formation of the complex. It is worth mentioning that the same complex was observed much earlier for Cd solute in Si or Ge [4].

The migration mechanism implying vacancies can then be described as follows. Denoting by the vectors $\{\omega_i\}$ (length $\omega$) the 1NN neighbours of the origin on which the OSA is located, let us assume that a vacancy V jumps from some site $R_{init}$ with $R_{init} \subset \{\omega_i + \omega_j\}$ belonging to more distant shells ($|R_{init}| > \omega$) towards a 1NN neighbour site of the OSA, say $\omega_{i_0}$ : then OSA slides without any activation barrier towards the intermediate site $\lambda_{i_0} = \omega_{i_0}/2$ while the vacancy splits into two halves located on $r = 0$ and $r = \omega_{i_0}$.

The migration process then goes on as follows :

*either the half-vacancy on $\omega_{i_0}$ jumps back towards one of its first neigbours $\omega_{i_0} + \omega_j$ (with $\omega_{i_0} + \omega_j \neq 0$) while rejecting simultaneously the OSA on the origin, in which case the net displacement of the OSA is equal to zero (Fig. 1b-2b);

*or the half-vacancy located on $r = 0$ jumps towards one of its first neighbours $\omega_{i_1}$ (with $\omega_{i_1} \neq \omega_{i_0}$) while rejecting simultaneously the OSA on site $\omega_{i_0}$, in which case the net displacement of the OSA is equal to $\omega_{i_0}$.

This picture holds for the BCC and FCC lattices. For the more densely packed FCC structure another possible migration path must be included: the pseudo-divacancy (made of two half-vacancies) can also migrate through a rotation without dissociating. The net displacement of the OSA takes place from an intermediate site to another, as depicted schematically in Fig. 3.

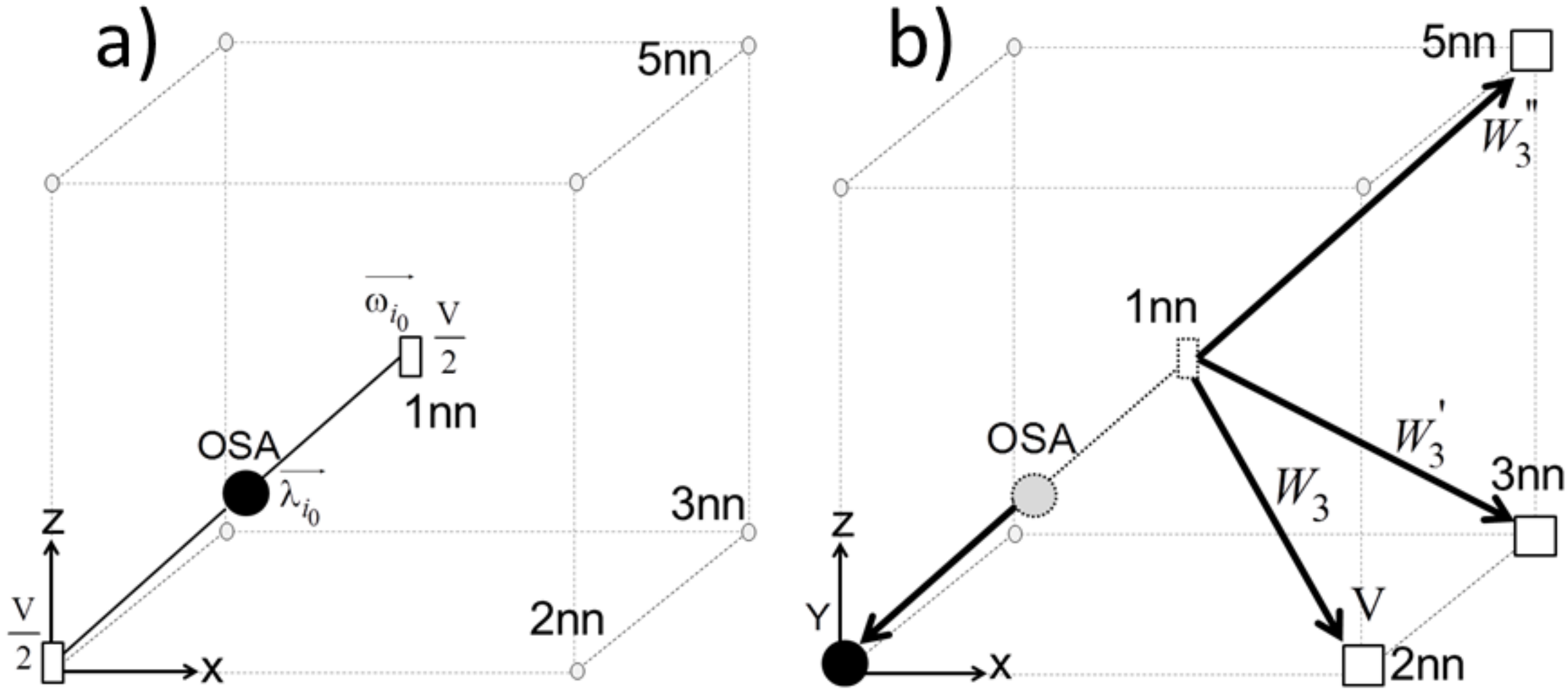

Figure 1. Dissociation of complex (V/2+OSA+V/2) through dissociative jumps in a BCC lattice: a) starting configuration; b) final configurations.

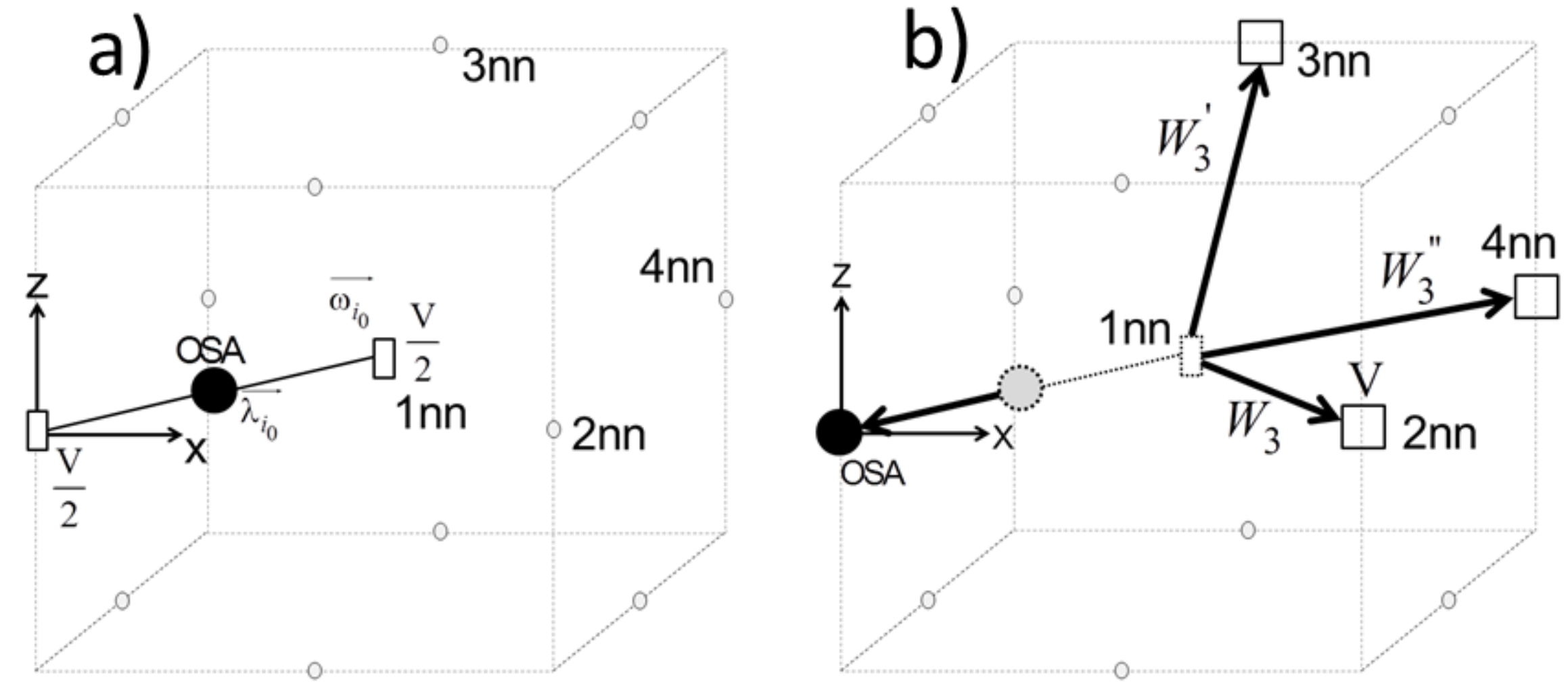

Figure 2. Dissociation of complex (V/2+OSA+V/2) through dissociative jumps in a FCC lattice: a) starting configuration; b) final configurations.

All the previous authors agreed that the standard expressions for the correlation factor could not be used. But for lack of anything better, former evaluations of Y diffusivity were made, either by neglecting the correlation effects, or by using existing models [2] to get some order of magnitude for the correlation effect (namely, Le Claire's model [5]).

The present contribution performs a thorough analysis of this variant of the vacancy mechanism and establishes the analytical expressions to be used for the solute diffusion coefficient $D_{B*}$ and for the exact value of the correlation factor $f_B$. Since much of the underlying formalism was already applied for the standard vacancy mechanism in previous publications [6-7], the present contribution will refer to them whenever necessary.

In a first part, we first reformulate the expression of the diffusion coefficient.

In a second part, we recall the main steps of the calculation, while keeping a general frame which is applied to a BCC lattice. The problem of correlation effects is shown to reduce to the solution of a system of linear equations, the unknowns of which are site occupancy probabilities of the vacancy in the neighbourhood of the tracer. We enlighten new results concerning the properties of the coefficients appearing in the equations. The particular case of Y in BCC iron is then studied, while adopting the first principle results obtained by previous authors for the solute-vacancy interactions up to the 5th neighbour, together with their calculation of migration barriers [2]. The variation with temperature of the diffusion coefficient and correlation factor is presented and the comparison with iron self-diffusion is made.

In a third part, the case of the FCC lattice is examined. In this part only the main differences with the BCC case will be quoted. The particular case of Y in FCC iron is then studied, while adopting the first-principles results obtained recently for yttrium-vacancy interactions up to the 7th neighbour, together with the migration barriers [12].

The conclusions are presented in the fourth part.

At last the details of the calculations, when too lengthy, are described in dedicated appendices.

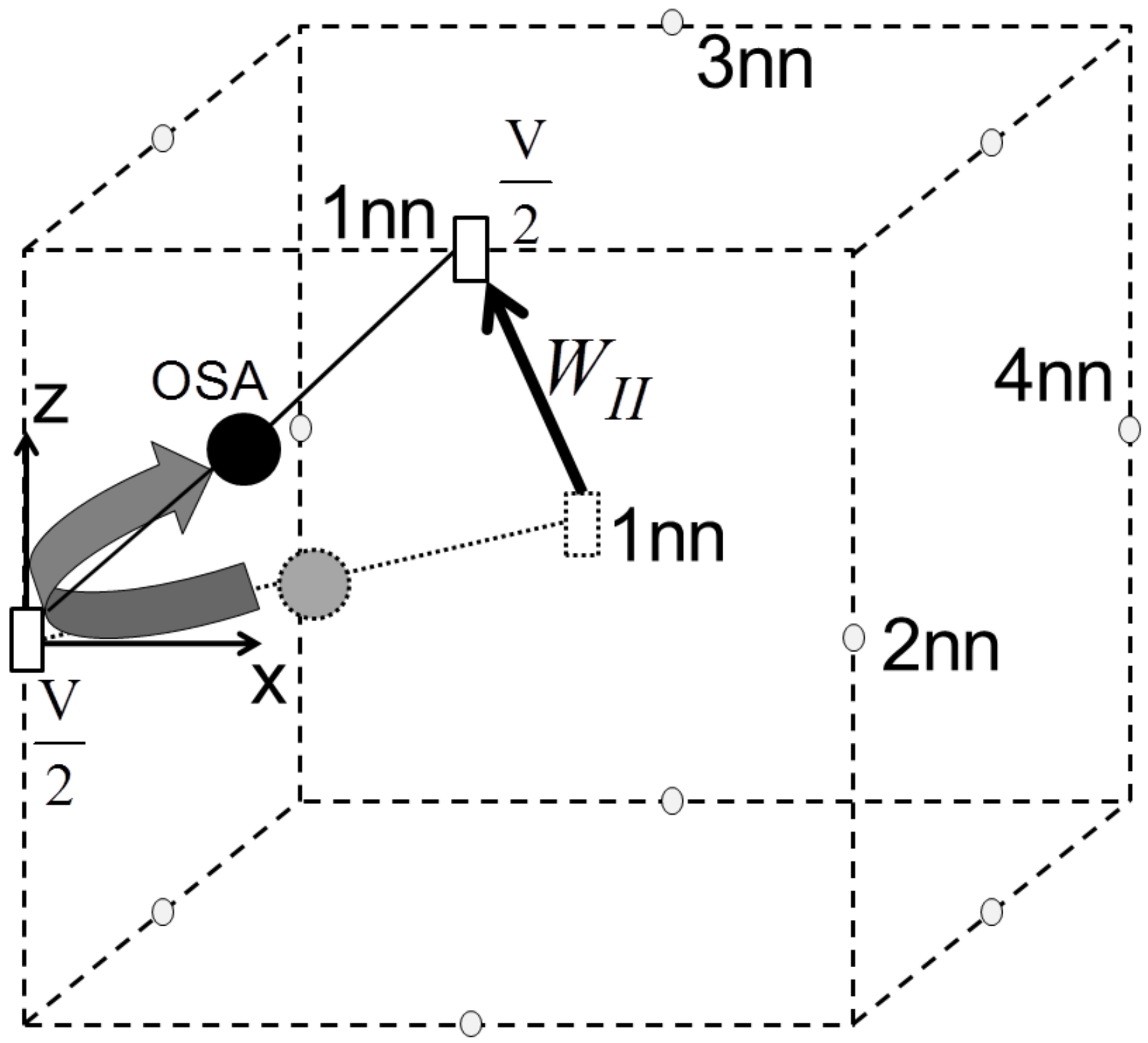

Figure 3. Non-dissociative jump of complex (V/2+OSA+V/2) in a FCC lattice. The meaning of the curved arrow is explained lower in section III.

## I New formulation of the tracer diffusivity

The diffusion coefficient of a tagged atom B* in infinite dilution is related to its average square displacement $< R^2(t) >$ during delay time ‘t’ by the Einstein formula $D_{B*} = \lim_{t\to\infty} \frac{< R^2(t) >}{6t}$. This displacement is the result of all the jumps performed with a large number of distinct defects over a large time interval.

These jumps can however be considered as:
* bunched in space: the number of jumps performed with only one vacancy is small (hardly larger than unity in 3D walks [8]) resulting in an overall displacement of a few lattice spacings only;
* bunched in time: the total vacancy concentration $C_{V0}$ is small and the time delay separating the arrival of two different vacancies on the tracer is large compared to the time spent by one given vacancy in its vicinity. This implies that, on the average, a defect labelled *k* will arrive in the neighbourhood of the tracer atom only a long time after the defect labelled *k*-1 definitely escaped after completion of its exchanges with the tracer.

The jumps performed with a given defect can thus be gathered together into what is called an encounter [9]. Strictly speaking, these encounters are not fully independent because of the finite vacancy concentration in the medium. If the OSA starts an encounter with a first vacancy, a second vacancy can approach the OSA before the first exhausts its probability of return; as a consequence there is some overlap between an encounter and the next one. For usual vacancy concentrations in metals and disordered alloys ($10^{-5}$ at% at most at the melting point) it has been shown that this overlap is totally negligible: the value of the correlation factor is changed by less than $10^{-6}$ [10]. This is the reason why the contributions of successive encounters will be assumed independent from one another and simply additive. The general formula above can thus be replaced by:

$$D_{B*} = \frac{< R^2 >_{Enc}}{6\Delta t_{Enc}} \tag{1}$$

where $< R^2 >_{Enc}$ is the average quadratic displacement of the tracer B* during one encounter with one vacancy and $\Delta t_{Enc}$ is the average duration allotted to an encounter or, in other terms, the average time delay which separates the successive arrivals of two distinct vacancies in the neighbourhood of the solute atom B*.

## II Diffusion coefficient in the BCC lattice

This section details the way how to calculate $< R^2 >_{Enc}$ and $\Delta t_{Enc}$.

### II-1 Calculating the average squared displacement during one encounter in the BCC lattice

The encounter starts at time zero, when the tracer atom B*, which was previously located on a substitutional site (denoted ‘S’), is pushed onto an intermediate location (denoted ‘I’), midway between two first-neighbour lattice sites ‘S’, by a vacancy which it never encountered before. The possible vectors for this S→I jump $\{\lambda_i\} = \{\omega_i / 2\}$ are collinear with the first neighbour vectors $\{\omega_i\}$. The length of $\{\lambda_i\}$ is denoted by $\lambda = \omega / 2$. Then the tracer atom B* comes back onto a substitutional location through an I→S jump of length $\lambda$, while expelling the vacancy on some neighbouring site. This set of elementary jumps (S→I + I→S) is called a macrojump; its total length can be equal to zero (if the initial and final substitutional sites coincide), or to $\omega$ (if they do not). After this first macrojump, the tracer can perform a second one with a probability $P$ (strictly smaller than unity for a 3D walk [8]), a third macrojump with a probability $P^2$ . . . etc. Finally the defect will escape to infinity or will be absorbed by a sink, which puts an end to the encounter.

The calculation of the corresponding quadratic average displacement of the tracer requires the introduction of probability functions attached to each type of jump and consists in writing recurrence relations for the displacement. Let be

$SI_n^{\lambda_i}(r)$ : the probability for B* atom of performing its n^th jump from lattice site $r$ to intermediate site $r+\lambda_i$ ;

$IS_n^{\lambda_j}(r+\lambda_i)$ : the probability for B* atom of performing its n^th jump from the intermediate site $r+\lambda_i$ to the lattice site $r+\lambda_i+\lambda_j$. For a BCC lattice, the allowed jump vectors $\lambda_j$ for this function reduce to $\pm\lambda_i$.

Simple recurrence equations for the displacement of the OSA can be established:

$$SI_n^{\lambda_i}(r) = \sum_{\{\lambda_j\}} IS_{n-1}^{-\lambda_i}(r+\lambda_j)\, p_{-\lambda_j,\lambda_i}$$
$$IS_n^{-\lambda_i}(r+\lambda_i) = \frac{1}{2}\left(SI_{n-1}^{\lambda_i}(r) + SI_{n-1}^{-\lambda_i}(r+\omega_i)\right) \qquad (2)$$

A detailed knowledge of these functions is not necessary, since the required quadratic displacement is related to the only knowledge of second order moments. Since diffusion is isotropic, an isotropic initial condition can be chosen, which reflects the fact that B* can be pushed initially onto an intermediate site in any of the z possible directions:

$$SI_1^{\lambda_i}(r) = \delta(r)/z \qquad (3)$$

Thanks to the spherical symmetry of the initial condition, the moments of IS and SI functions have a magnitude which is independent of the direction indicated by their superscript. Zeroth, first and second order moments of IS or SI functions represent the number of jumps, the total displacement and the squared displacement in each direction respectively, which correspond to I→S or S→I jumps. First order moments are vectors which sum up to zero since there is presently no drift force. Zeroth, first and second order moments of these functions are thus defined as follows:

$$\sum_{n=1}^{\infty}\sum_{\{r\}} SI_n^{\lambda_i}(r) = M_{SI0}$$
$$\sum_{n=1}^{\infty}\sum_{\{r\}} (r+\lambda_i) SI_n^{\lambda_i}(r) = M_{SI1}\overrightarrow{\lambda_i}$$
$$\sum_{n=1}^{\infty}\sum_{\{r\}} (r+\lambda_i)^2 SI_n^{\lambda_i}(r) = M_{SI2}\lambda^2$$
$$\sum_{n=1}^{\infty}\sum_{\{r\}} IS_n^{-\lambda_i}(r+\lambda_i) = M_{IS0} \qquad (4)$$
$$\sum_{n=1}^{\infty}\sum_{\{r\}} r IS_n^{-\lambda_i}(r+\lambda_i) = -M_{IS1}\overrightarrow{\lambda_i}$$
$$\sum_{n=1}^{\infty}\sum_{\{r\}} r^2 IS_n^{-\lambda_i}(r+\lambda_i) = M_{IS2}\lambda^2$$

The sites on which the moments are evaluated are the arrival sites of the corresponding functions, i.e. the site obtained by summing the argument and the superscript. The detailed calculation of these moments is reported in Appendix A.

The average number of macrojumps is given by $\left(1-P\right)^{-1}$.The mean square displacement corresponding to the encounter with a given vacancy is expressed with the second order moment of those functions which bring the tracer atom back onto a substitutional site, i.e. the IS ones. Hence:

$$< R^2 >_{Enc} = \frac{\sum\limits_{\{\lambda_i\}} \sum\limits_{n=1}^{\infty} \sum\limits_{\{r\}} r^2 IS_n^{-\lambda_i}(r+\lambda_i)}{\sum\limits_{\{\lambda_i\}} \sum\limits_{n=1}^{\infty} \sum\limits_{\{r\}} IS_n^{-\lambda_i}(r+\lambda_i)} = \frac{zM_{IS2}\lambda^2}{zM_{IS0}} = \frac{2\lambda^2(1+Q)}{(1-P)}, \tag{5}$$

where $P$ is the total probability of performing a S→I jump after a I→S one with the same vacancy, $Q$ is the average cosine between an I→S jump and the next S→I one. The mean square length of a macrojump is given by the above expression after setting $P = Q = 0$, i.e. when the tracer is allowed to perform only one macrojump. Hence $< R^2 >_{MJ} = 2\lambda^2 = \omega^2 / 2$.

The random square displacement produced by $\left(1-P\right)^{-1}$ macrojumps is then :

$$< R^2 >_{rand} = \left(1-P\right)^{-1} 2\lambda^2$$

and the correlation factor is by definition given by the ratio :

$$f_B = \frac{< R^2 >_{Enc}}{< R^2 >_{rand}} = 1+Q\,. \tag{6}$$

In this transport mechanism, the displacements making up a given macrojump are not correlated; the only correlation effect takes place between the I→S jump of macrojump number ‘n’ and the S→I jump of macrojump number ‘n+1’. A similar situation was already encountered in the past for the interstitialcy mechanism in ionic crystals [5,10].

### II-2 Calculating the time delay allotted to an encounter in the BCC lattice

We define the reference state of the energy as a crystal containing a tracer atom B* on the origin and a vacancy far apart from it in the bulk. The vacancy concentration in the bulk is $C_{V0}$ and, at closer distances, $C_{Vi} = C_{V0} e^{-\beta E_i}$ where $E_i$ is the vacancy-solute interaction energy when the vacancy sits on the i$^{th}$ neighbour shell of the tracer ($E_i < 0$ for an attraction and $E_i > 0$ for a repulsion).

The average time delay which is necessary to perform a macrojump must first be calculated. For the transport mechanism under study, the average frequency $\Gamma_{SI}$ of an S→I jump is the total jump frequency of a vacancy towards a first neighbour site of the tracer. The sites $R_{init}$ it starts from belong to more distant shells (labelled ‘j’) than the first one and the vacancy jump frequencies from shell ‘j’ to shell ‘1’ are named $W_{j\to 1}^{shell}$ (j= 2, 3, 5 for the BCC lattice). The probability of finding a vacancy on shell ‘j’ is by definition its atomic concentration $C_{Vj}$. Denoting by $nbond_{1\to j}$ the number of bonds connecting a given site of the 1$^{st}$ shell to sites of the j$^{th}$ shell, the frequency $\Gamma_{SI}$ is given by:

$$\Gamma_{SI} = z\sum_{jV1} nbond_{1\to j} C_{Vj} W_{j\to 1}^{shell} = zC_{V0}\sum_{jV1} nbond_{1\to j} e^{-\beta E_j} W_{j\to 1}^{shell} \tag{7}$$

where
* $z$ is the number of first neighbours;
* the summation $\sum_{jV1}$ runs on the shells which can be reached from the first shell with one jump.

The frequency $\Gamma_{IS}$ of an I→S jump is given by:

$$\Gamma_{IS} = 2\sum_{jV1} nbond_{1\to j} W_{1\to j}^{shell} \tag{8}$$

where the multiplicative factor equal to 2 takes into account the two possibilities for returning on a substitutionnal site.

Since the jumps S→I + I→S are performed in series, their delays are additive :

$$\Delta t_{MJ} = \left(\Gamma_{SI}\right)^{-1} + \left(\Gamma_{IS}\right)^{-1}$$

and the frequency of a macrojump is defined as

$$\Gamma_{MJ} = (\Delta t_{MJ})^{-1} = \Gamma_{SI}\Gamma_{IS} / \left(\Gamma_{SI} + \Gamma_{IS}\right) \tag{9}$$

The duration of an encounter made of $\left(1-P\right)^{-1}$ macrojumps is then $\Delta t_{Enc} = \left(1-P\right)^{-1} \Delta t_{MJ}$ and the tracer diffusion coefficient is finally expressed as:

$$D_B^* = \frac{< R^2 >_{Enc}}{6\Delta t_{Enc}} = \frac{\Gamma_{SI}\Gamma_{IS}}{12(\Gamma_{SI} + \Gamma_{IS})} \omega^2 (1+Q)\,. \tag{10}$$

**II-3 Calculation of Q in the BCC lattice**

The method consists in writing down a transport equation for the function $L(r,t)$ which stands for the probability of finding the vacancy on site $r$ at time t. The initial condition $L(r,o) = CI(r)$ stores the position $R_{init}$ of the vacancy after an I→S jump has carried back the tracer onto a lattice site, which is taken as the origin later on. The initial condition once given, the correlation problem reduces to the calculations of the return probabilities of the vacancy onto the tracer atom to let it perform the next jump. It has been shown at length that these return probabilities are proportional to the time cumulated occupancies on those sites $R_{init}$ from which the vacancy can induce a further macrojump: they are given by the Laplace transform $LL(r,p)$ of $L(r,t)$, when evaluated on these sites [6-7]. At last the Fourier transform takes into account the whole extent of the lattice beyond the first shells around the tracer atom. Thus, the general equation will have to be Laplace and Fourier transformed.

Far from B*, i.e. in the bulk, the time derivative of this probability $L(r,t)$ is given by a simple balance on site $r$ between ingoing and outgoing contributions:

$$\frac{dL(r,t)}{dt} = -zW_O L(r,t) + W_O \sum_{\{\omega_i\}} \left[L(r+\omega_i, t)\right], \tag{11}$$

where $W_O$ stands for the vacancy jump frequency in the bulk.

This general expression must then be corrected in order to take into account *specific sites*, i.e. sites with properties departing from those of a bulk site: for a specific site, some (or all) of the outgoing or ingoing frequencies depart from $W_O$. These *specific* *sites* are denoted by $R_j$ (j =1,N). For each specific site, a $\delta()$ operator is used to add a corrective term which modifies the general balance written above. The complete transport equation reads now:

$$\frac{dL(r,t)}{dt} = -zW_O L(r,t) + W_O \sum_{\{\omega_i\}} \left[ L(r+\omega_i,t) \right] + \\ \sum_{\{R_j\}} \delta(r-R_j) \sum_{\{\omega_i\}} \delta(R_j+\omega_i-\{R_k\}) \left[ L(R_j,t)(W_O - W_{R_j \to R_k}) + L(R_j+\omega_i,t)(W_{R_k \to R_j} - W_O) \right] \quad (12)$$

The second line includes all the corrections for specific sites and uses two summations in series:

* the first summation scans all the specific sites;
* the second summation runs over the first neighbours of the specific site $R_j$ : if some vector $R_j + \omega_i$ points at any of specific sites $R_k$, the outgoing (ingoing) frequency from (to) site $R_j$ is different from $W_O$ and must be replaced by $W_{R_j \to R_k}$ ($W_{R_k \to R_j}$) which stands for the jump frequency from the specific site $R_j$ ($R_k$) to the specific site $R_k$ ($R_j$). This frequency is set to zero if $R_j$ and $R_k$ are not first neighbours or if the starting site is a sink i.e. a site which the vacancy cannot escape from. Due to the constraint imposed by the two $\delta()$ operators in series, the internal summation runs only on those sites for which at least one of the outgoing or ingoing frequency is different from $W_O$. The doubly transformed function is then:

$$FLL(k,p) = \frac{FCI(k)}{Denom} + \\ \sum_{\{R_j\}} \frac{e^{-ikR_j}}{Denom} \sum_{\{\omega_i\}} \delta(R_j+\omega_i-\{R_k\}) \left( -LL_j (W_{R_j \to R_k} - W_O) + LL_k (W_{R_k \to R_j} - W_O) \right), \quad (13)$$

where:

* $FLL(k,p)$ : the Fourier Laplace transform of $L(r,t)$ ;
* $FCI(k)$ : the Fourier transform of the initial condition;
* $Denom = p + W_O D_0$ with $D_0 = z - \sum_{\{\omega_i\}} e^{-ik\omega_i}$
* $LL_j = LL(R_j,p) = \int_0^\infty L(R_j,t) e^{-pt} dt$ : the Laplace transform of $L(r,t)$ evaluated on site $R_j$.

In most cases of interest, the interaction between tracer B* and the vacancy is isotropic since it depends only on their mutual distance: the jump frequency $W_{R_j \to R_k}$ does not depend on the sites $R_j$ and $R_k$ but only on the neighbour shells $R_j$ and $R_k$ belong to; the number of distinct frequencies is appreciably reduced and the symbol $W_{j \to k}^{shell}$ now stands for the vacancy jump frequency from a site belonging to shell 'j' to a site belonging to shell 'k'. Going on further, it was shown previously [6-7] that taking into account the symmetry of the lattice and adopting an initial condition with a mirror antisymmetry reduced the dimension of the problem. The specific sites which are occupied with an equal probability (in absolute value) by the wandering vacancy can be grouped into $M_{sub}$ subsets ($M_{sub} << N$). Each subset 'j' consists of $n_j^+$ sites $\left\{R_j^+\right\}$ on the positive side of the x-axis which are visited with a cumulative probability $LL_j^+ > 0$ and $n_j^- = n_j^+$ mirror symmetry sites $\left\{R_j^-\right\} = -\left\{R_j^+\right\}$ on the negative side of the x-axis, which are visited with a time cumulated probability $LL_j^- = -LL_j^+$. Lattice sites $\left\{R_j\right\} = \left\{R_j^+\right\} \cup \left\{R_j^-\right\}$ belong to the same neighbour shell. At last subsets $0 \le j \le m_{sub} - 1$ are contained in the mirror plane x=0: their introduction is necessary for defining a complete set of jump frequencies, but they play no role afterwards since they have a zero occupancy probability thanks to the mirror antisymmetry of the initial condition. The doubly transformed function can then be reduced to a summation with an index 'j' running no longer on specific sites but only on subsets:

$$FLL(k,p) = \frac{FCI(k)}{Denom} + \sum_{j=m_{sub}}^{M_{sub}} \frac{W_O f_j}{Denom} \sum_{kVj} nlink_{j \to k} \left[ -LL_j^+ \; W_{j \to k}^{'sub} + LL_k^+ \; W_{k \to j}^{'sub} \right], \tag{14}$$

where:

* $M_{sub}$ : the total number of subsets;
* $m_{sub}$ : the minimum value of the index restricting the summation to relevant subsets;
* $f_j = \sum_{\left\{R_j^+\right\}} e^{-ikR_j^+} - \sum_{\left\{R_j^-\right\}} e^{-ikR_j^-} = -2i \sum_{\left\{R_j^+\right\}} sin(kR_j^+)$ : a function attached to the set of specific sites $\left\{R_j^+\right\} \cup \left\{R_j^-\right\}$;
* $\sum_{kVj}$ : the neighbourhood between lattice sites is transposed into a neighbourhood between subsets; the summation is restricted to subsets 'k' which contain at least one site which is a first neighbour of at least one site belonging to subset 'j';
* $nlink_{j \to k}$ : the number of sites of subset 'k' which can be reached in one jump from a given site of subset 'j'; the value is set to zero whenever subset 'k' is not a first neighbour of subset 'j'; they obey the obvious condition $\sum_{kVj} nlink_{j \to k} = z$; Appendix B gives in Tables B1-B2 the values of $nlink_{j \to k}$ for BCC and FCC lattices which were extracted from the standard output of the code supplemented with [7].
* $LL_j^+ = LL(R_j^+, p) = -\frac{1}{n_j V_{ZB}} \int_{V_{ZB}} f_j \; FLL\,(k,p)\; d_3k$ : the Laplace transform of $L(r,t)$ calculated on sites $\left\{R_j^+\right\}$ with $n_j = n_j^+ + n_j^-$ ;
* $W_{j \to k}^{'sub} = (W_{j \to k}^{sub} - W_O)/W_O$ : the relative jump frequency connecting subsets 'j' and 'k'.

It is worth mentioning that 'j' and 'k' stand now for indexes of subsets and no longer of shells as above. The knowledge of jump frequencies between shells, which is based on physical interactions, can be easily transposed to jump frequencies between subsets, the latter implying additional symmetry considerations. In order to avoid errors, an automatic scanning of neighbour shells is necessary to establish the link between the index 'j' of a subset and the index of the neighbour shell subset 'j' belongs to; this is one of the tasks which are carried out by the computer program which was given as supplemental material [7].

It is worth noticing that, up to this point, $LL_j^+$ still depends on the Laplace variable $p$. The interesting quantity for the correlation problem is the time cumulated probability, that is, the limit when $p \to 0$. We will denote with the same symbol the Laplace transform and its limit when $p \to 0$.

**II-4 Establishing the equations of the linear system**

We first establish a more convenient expression of the doubly transformed equation: it is obtained after gathering together all the terms corresponding to the same $LL_j^+$ in Eq. (14), in order to recast it under the form:

$$FLL(k,p) = \frac{FCI(k)}{Denom} + \sum_{j=m_{sub}}^{M_{sub}} coef(j)\, LL_j^+ . \tag{15}$$

The detailed derivation of the expressions defining the coefficients $coef(j)$ is given in Appendix C:

$$coef(j) = -\frac{W_O}{Denom} \sum_{kVj} W'^{sub}_{j\to k} \left( f_j\, nlink_{j\to k} - f_k\, nlink_{k\to j} \right). \tag{16}$$

Obtaining the system of linear equations to determine the unknowns $LL_i^+$ is then straightforward with Eq. (15-16). Taking the reverse Fourier transform for sites $\{R_i\}$ yields the equation number ‘i’ which is used to define $LL_i^+$ according to:

$$\begin{aligned} LL_i^+ &= -\frac{1}{n_i V_{ZB}} \int_{V_{ZB}} f_i\, FLL(k,p) d_3k \\ &= -\frac{1}{n_i V_{ZB}} \int_{V_{ZB}} \frac{f_i\, FCI(k)}{Denom} d_3k \\ &+ \sum_{j=m_{sub}}^{M_{sub}} LL_j^+ \frac{1}{n_i V_{ZB}} \int_{V_{ZB}} \frac{W_O f_i \sum_{kVj} W'^{sub}_{j\to k} \left( f_j\, nlink_{j\to k} - f_k\, nlink_{k\to j} \right)}{Denom} d_3k \end{aligned}$$

The system is then recast into the general form

$$\sum_{j=m_{sub}}^{M_{sub}} M_{ij}\, LL_j^+ = RHS_i \qquad i \in [m_{sub}, M_{sub}] \tag{17}$$

with:

$$* RHS_i = -\frac{1}{n_i V_{ZB}} \int_{V_{ZB}} \frac{f_i\, FCI(k)}{p + W_O D_0} d_3k$$

$$* M_{ij} = \delta(i-j) + \sum_{kVj} W'^{sub}_{j\to k} \left( fifj(i,j)\, nlink_{j\to k} - fifj(i,k)\, nlink_{k\to j} \right)$$

$$* fifj(i,j) = -\frac{1}{n_i V_{ZB}} \int_{V_{ZB}} \frac{W_0 f_i f_j}{p + W_0 D_0} d_3k$$

The knowledge of all these probabilities gives an access to the evaluation of $Q^{BCC}$. In the present study, where the numbering of subsets corresponds to solute-vacancy interactions extending up to the 5th neighbour shell, the average cosine is given by:

$$Q^{BCC} = -\sum_{jV5} n_j^+ LL_j^+ (nlink_{j\to 5} W^{sub}_{j\to 5}) \tag{18}$$

It should be kept in mind that the lattice integrals $fifj(i,j)$ still depend on the Laplace variable. The only interesting value is their limit $-\frac{1}{n_i V_{ZB}}\int_{V_{ZB}} \frac{f_i f_j}{D_0} d_3k$ which is obtained when $p \to 0$: this limit will be denoted by the same symbol.
The limit for $p \to 0$ can be taken for all equations but those associated with a subset containing sink sites, i.e. sites, the vacancy cannot escape from. Indeed, if subset 'j' is made of such sink sites, the quantity $LL_j^+$ diverges like $p^{-1}$ and cannot be manipulated like the other unknowns. A modified expression of the coefficient $coef(j)$ is needed as explained below.

### II-5 Linear equation system: properties of coefficients in special cases

One particular case of interest deserves special attention. When the escape frequencies of the vacancy depend only on the starting subset 'j' and not on the subset 'k' toward which it jumps, i.e. $W'^{sub}_{j\to k} = W'^{sub}_{j\to out}$, then the coefficients above become:

$$M_{ij} = \delta(i-j) - W'^{sub}_{j\to out}\sum_{kVj} \frac{1}{n_i V_{ZB}}\int_{V_{ZB}} \frac{W_0 f_i \left( f_j \; nlink_{j\to k} - f_k \; nlink_{k\to j}\right)}{p+W_0 D_0} d_3k$$

$$= \delta(i-j) - W'^{sub}_{j\to out} \frac{1}{n_i V_{ZB}}\int_{V_{ZB}} \frac{W_0 f_i (zf_j - \sum_{kVj} f_k \; nlink_{k\to j})}{p+W_0 D_0} d_3k$$

In Appendix D a proof is brought for the identity $zf_j - \sum_{kVj} nlink_{k\to j} \, f_k = f_j D_0$ and for the orthogonality of functions $f_i$ with the result $-\frac{1}{n_i V_{ZB}}\int_{V_{ZB}} f_i f_j d_3k = \delta(i-j)$.
Multiplying both sides of the identity by $f_i / D_0$ and integrating over the first Brillouin zone gives the following identities:

$$z \; fifj(i,j) - \sum_{kVj} nlink_{k\to j} \, fifj(i,k) = \delta(i-j) \tag{19}$$

When formulated in a universal way which is independent from the way of numbering the subsets (i.e. replacing the number of the subset by the coordinates of its representative point), the relationships become:

$$z \; fifj_{i_1 i_2 i_3 \times j_1 j_2 j_3} - \sum_{kVj} nlink_{k\to j} \, fifj_{i_1 i_2 i_3 \times k_1 k_2 k_3} = \delta_{i_1 i_2 i_3 - j_1 j_2 j_3}. \tag{20}$$

where:
* $(i_1,i_2,i_3)$ and $(j_1,j_2,j_3)$ are the coordinates of the representative site for subset 'i' and 'j' respectively; this representative site belongs to the first octant and comply with $(i_1 \geq 0,\; 0 \leq i_2 \leq i_3)$ and $(j_1 \geq 0,\; 0 \leq j_2 \leq j_3)$ respectively;
* $fifj(i,j) = fifj_{i_1 i_2 i_3 \times j_1 j_2 j_3}$ ;
* $\delta j_{i_1 i_2 i_3 - j_1 j_2 j_3} = \delta(i_1 - j_1)\delta(i_2 - j_2)\delta(i_3 - j_3)$.

This formal result establishes in a general way the relationships between lattice integrals which were noticed previously and proved on some particular cases [7].

Such identities can be established as long as all the neighbours 'k' of subset 'j' are explicitly introduced in the calculation. As an illustration, we give in Table D1 of Appendix D the identities obtained for a solute-vacancy interaction range extending up to the 5th and 7th neighbour shell in the BCC and FCC lattices, respectively.

Whenever the escape frequency does not depend on the sites the vacancy jumps to, the coefficients thus become:

$$\begin{aligned} M_{ij} &= \delta(i-j) - W'^{sub}_{j\to out} \frac{1}{n_i V_{ZB}} \int_{V_{ZB}} \frac{W_0 f_i f_j D_0}{p + W_0 D_0} d_3k \\ &= (1 + W'^{sub}_{j\to out})\delta(i-j) + W'^{sub}_{j\to out} \frac{p}{n_i V_{ZB}} \int_{V_{ZB}} \frac{f_i f_j}{p + W_0 D_0} d_3k \end{aligned} \quad (21)$$

This result leads to the further observations :

*off-diagonal coefficients vanish when $p \to 0$. In particular, if all the subsets 'j' which are neighbours of 'subset 'i' have a uniform escape frequency, then $LL_i^+$ is no longer coupled to these $LL_j^+$ in the equations. This implies that the correlation factor can become independent of the interaction energies at some sites from which the escape frequencies are all equal (cf. case of Model-I in [7]).

*diagonal coefficients vanish too if subset 'j' is a sink ($W'^{sub}_{j\to out} = -1$) on which return probability accumulates. In this case $LL_j^+ \to \infty$ when $p \to 0$ and cannot be kept as is in the system of unknowns. This is the reason why in the original formulation of the method, such a quantity was analytically eliminated at the benefit of the other unknowns [10]. In a recent application to the vacancy mechanism, the only diverging quantity was attached to the origin, i.e. to a site belonging to the mirror antisymmetry plane which disappeared from the system of equations [7]. It is no longer the case presently and a dedicated treatment of the corresponding term must be performed. Eq. (21) contains the remedy to this problem thanks to the presence of a first order term in $p$ which counterbalances the divergence: it is seen that the quantity $LL_j^+$ is thus naturally replaced by the product $p\, LL_j^+ / W_0$ which remains bounded and plays the role of a new variable: this variable can now be formally manipulated in the same way as the others. The corresponding coefficient is given by Eq. (21) after $p \to 0$ and is expressed by $M_{ij} = -\frac{1}{n_i V_{ZB}} \int_{V_{ZB}} \frac{f_i f_j}{D_0} d_3k = fifj(i,j)$.

## II-6 Application to the case of Y in a BCC lattice

Let us define: $W_{IS} = 3W^{shell}_{1\to 2} + 3W^{shell}_{1\to 3} + W^{shell}_{1\to 5}$ together with the ratios

$$c_3 = \frac{W^{shell}_{1\to 2}}{W_{IS}} \qquad c'_3 = \frac{W^{shell}_{1\to 3}}{W_{IS}} \qquad c''_3 = \frac{W^{shell}_{1\to 5}}{W_{IS}} \text{ with } 3c_3 + 3c'_3 + c''_3 = 1 \quad (22)$$

An initial condition complying with the four-fold symmetry around <100> is introduced. The vacancy separating from the B* atom through a dissociative jump from site $r = \omega_{111}$ is replaced by four quarters of vacancy starting from $r = \omega_{111}, \omega_{1\bar{1}1}, \omega_{11\bar{1}}, \omega_{1\bar{1}\bar{1}}$. A negative source is introduced along the negative x-axis on sites $r = \omega_{\bar{1}11}, \omega_{\bar{1}\bar{1}1}, \omega_{\bar{1}1\bar{1}}, \omega_{\bar{1}\bar{1}\bar{1}}$ by mirror antisymmetry. The initial condition of the diffusion problem is given by:

$$\begin{aligned}
CI(r,0) = &\, \mathit{4}\ \frac{c_3}{4}\delta(r-\omega_{200}) + \mathit{2}\ \frac{c_3^{'}}{4}\left(\delta(r-\omega_{202})+\delta(r-\omega_{20\bar{2}})+\delta(r-\omega_{220})+\delta(r-\omega_{2\bar{2}0})\right) \\
&+\mathit{1}\ \frac{c_3^{''}}{4}\left(\delta(r-\omega_{222})+\delta(r-\omega_{2\bar{2}2})+\delta(r-\omega_{22\bar{2}})+\delta(r-\omega_{2\bar{2}\bar{2}})\right) \\
&-\mathit{4}\ \frac{c_3}{4}\delta(r-\omega_{\bar{2}00}) - \mathit{2}\ \frac{c_3^{'}}{4}\left(\delta(r-\omega_{\bar{2}02})+\delta(r-\omega_{\bar{2}0\bar{2}})+\delta(r-\omega_{\bar{2}20})+\delta(r-\omega_{\bar{2}\bar{2}0})\right) \\
&-\mathit{1}\ \frac{c_3^{''}}{4}\left(\delta(r-\omega_{\bar{2}22})+\delta(r-\omega_{\bar{2}\bar{2}2})+\delta(r-\omega_{\bar{2}2\bar{2}})+\delta(r-\omega_{\bar{2}\bar{2}\bar{2}})\right)
\end{aligned}$$

where the italic digit reflects the number of those sites of the first shell which contribute to the presence of the vacancy on a given site of the 2nd, 3rd or 5th neighbour shell. The contributions coming from plane x=0 cancel out each other thanks to the antisymmetry mirror plane and disappear from the formula. The Fourier transform of initial condition becomes:

$$FCI(k,0) = \frac{c_3}{4} nlink_{6\to5} f_6 + \frac{c_3^{'}}{4} nlink_{7\to5} f_7 + \frac{c_3^{''}}{4} nlink_{10\to5} f_{10} \tag{23}$$

where $s_{jx} = \sin(jk_x)$ and $c_{jx} = \cos(jk_x)$.

The value of the average cosine will then be equal to:

$$Q^{BCC} = -n_6^+ LL_6^+ (nlink_{6\to5} W_{6\to5}^{sub}) - n_7^+ LL_7^+ (nlink_{7\to5} W_{7\to5}^{sub}) - n_{10}^+ LL_{10}^+ (nlink_{10\to5} W_{10\to5}^{sub}) \tag{24}$$

In the BCC lattice $nlink_{6\to5} = 4,\ nlink_{7\to5} = 2,\ nlink_{10\to5} = 1,\ n_6^+ = 1,\ n_7^+ = 4,\ n_{10}^+ = 4$ and $W_{6\to5}^{sub} = W_{2\to1}^{shell} = W_4$, $W_{7\to5}^{sub} = W_{3\to1}^{shell} = W_4^{'}$, $W_{10\to5}^{sub} = W_{5\to1}^{shell} = W_4^{''}$, which leads to the final expressions:

$$FCI(k,0) = c_3 f_6 + \frac{c_3^{'}}{2} f_7 + \frac{c_3^{''}}{4} f_{10},$$

$$Q^{BCC} = -4LL_6^+ W_4 \quad -8LL_7^+ W_4^{'} \quad -4LL_{10}^+ W_4^{''}.$$

Using the same standard notations, with $nbond_{1\to2} = 3,\ nbond_{1\to3} = 3,\ nbond_{1\to5} = 1$ yields the expressions of the frequencies entering the definition of the macrojump frequency:

$$\begin{aligned}
\Gamma_{SI} &= 8C_{V0}\left(3e^{-\beta E_2}W_4 + 3e^{-\beta E_3}W_4^{'} + e^{-\beta E_5}W_4^{''}\right) \\
\Gamma_{IS} &= 2\left(3W_3 + 3W_3^{'} + W_3^{''}\right)
\end{aligned} \tag{25}$$

The diffusion coefficient is finally expressed as:

$$\begin{aligned}
D_B^* &= \frac{< R^2 >_{Enc}}{6\Delta t_{Enc}} = \frac{\Gamma_{SI}\Gamma_{IS}}{6(\Gamma_{SI}+\Gamma_{IS})}\frac{\omega^2}{2}(1+Q^{BCC}) \\
&= \frac{2}{3}\frac{C_{V0}\left(3e^{-\beta E_2}W_4 + 3e^{-\beta E_3}W_4^{'} + e^{-\beta E_5}W_4^{''}\right)\left(3W_3 + 3W_3^{'} + W_3^{''}\right)}{4C_{V0}\left(3e^{-\beta E_2}W_4 + 3e^{-\beta E_3}W_4^{'} + e^{-\beta E_5}W_4^{''}\right) + \left(3W_3 + 3W_3^{'} + W_3^{''}\right)}\omega^2 f_B
\end{aligned} \tag{26}$$

For Y in BCC iron, the values of the interaction energies and of the saddle configurations are gathered below in the diagram of Fig. 4. Empty circles correspond to the energies $E_1$, $E_2$, $E_3$ … the Y-vacancy configurations, with a vacancy on a 1rst, $2^{nd}$, $3^{rd}$ … neighbour shell of the yttrium; interaction energies are written in black. Lines denote a first neighbour jump connection between two configurations. The small empty bold circle corresponds to the saddle position midway the two stable adjacent ones: the energy is reported in red, when the saddle energy was explicitly evaluated, and in blue when it was assumed equal to the bulk one for simplicity. The migration barrier is obtained by taking the energy of the saddle configuration minus the energy of the starting configuration.

A vacancy formation energy and entropy, equal to 2.12 eV and 4.08 $k_B$ respectively, were used. The pre-exponential term is taken equal to the Debye frequency $10^{13}$ $s^{-1}$ for all jump frequencies. The vacancy migration energy in the bulk is found equal to 0.69 eV, and the lattice parameter is found equal to 2.87 $10^{-10}$ m [2].

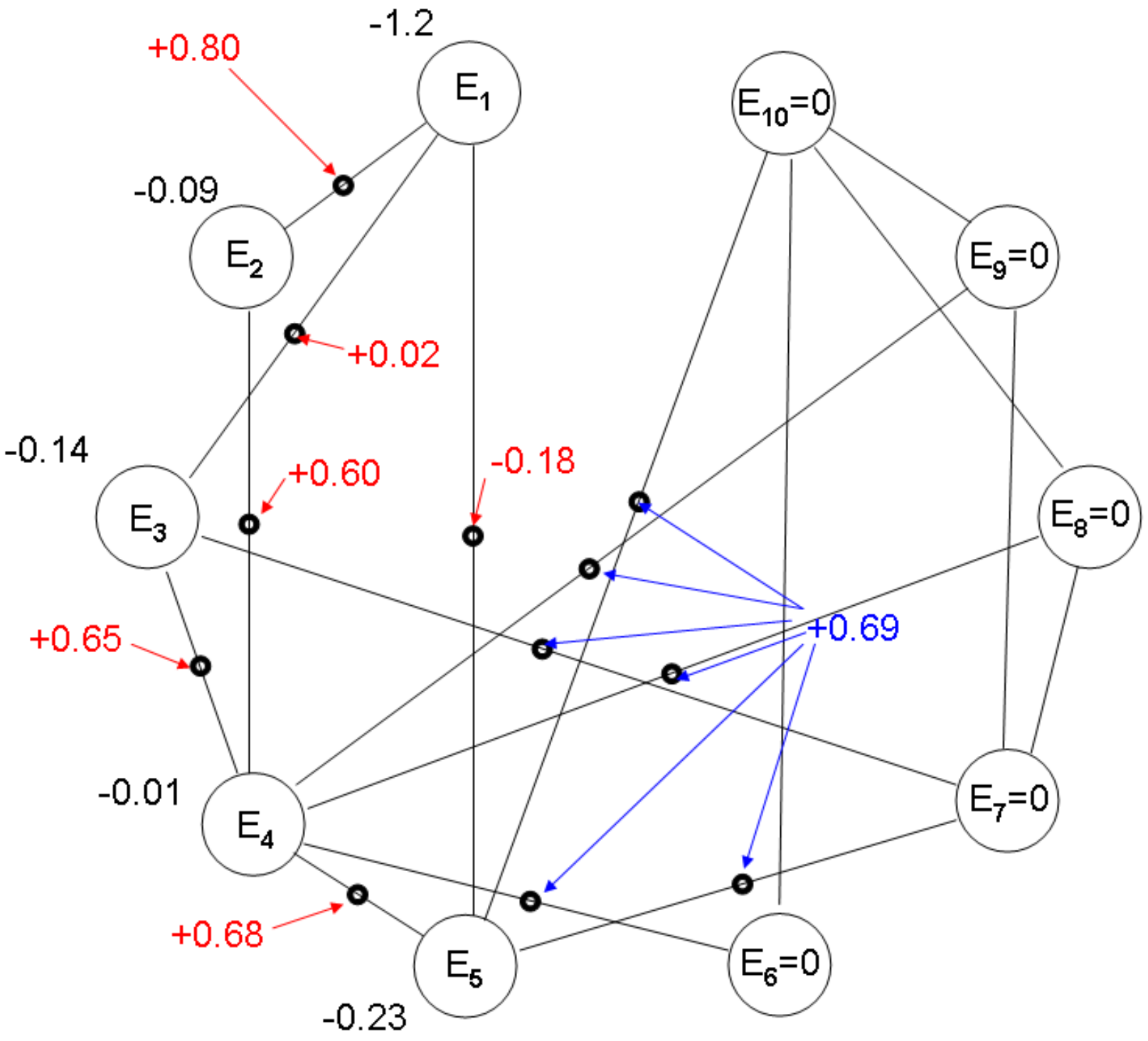

Figure 4. Diagram of interaction and saddle configuration energies for Y in BCC iron (eV). Data in black are the Y-V interactions at rest; data in red are the calculated saddle energies for the migration from $E_i$ to $E_j$ [2]; data in blue are the saddle energies for the vacancy migration in the bulk.

| T(K) | $f_Y^{exact}$ | $\Gamma_{MJ}^{exact}$ | $D_{Y*}^{exact}$ | $D_{Y*}^{Barouh}$ | $D_{Fe*}$ |
|---|---|---|---|---|---|
| 300 | 4.356 $10^{-4}$ | 1.213 $10^{-17}$ | 2.723 $10^{-41}$ | 1.576 $10^{-41}$ | 2.180 $10^{-52}$ |
| 320 | 7.065 $10^{-4}$ | 1.322 $10^{-15}$ | 4.809 $10^{-39}$ | 2.526 $10^{-39}$ | 1.945 $10^{-49}$ |
| 340 | 1.081 $10^{-3}$ | 8.301 $10^{-14}$ | 4.621 $10^{-37}$ | 2.229 $10^{-37}$ | 7.803 $10^{-46}$ |
| 370 | 1.876 $10^{-3}$ | 1.786 $10^{-11}$ | 1.725 $10^{-34}$ | 7.448 $10^{-35}$ | 1.860 $10^{-43}$ |
| 400 | 2.993 $10^{-3}$ | 1.719 $10^{-9}$ | 2.649 $10^{-32}$ | 1.041 $10^{-32}$ | 1.381 $10^{-40}$ |
| 440 | 5.040 $10^{-3}$ | 2.885 $10^{-7}$ | 7.486 $10^{-30}$ | 2.648 $10^{-30}$ | 2.286 $10^{-37}$ |
| 500 | 9.368 $10^{-3}$ | 1.356 $10^{-4}$ | 6.542 $10^{-27}$ | 2.038 $10^{-27}$ | 1.665 $10^{-33}$ |
| 540 | 1.306 $10^{-2}$ | 3.854 $10^{-3}$ | 2.592 $10^{-25}$ | 7.539 $10^{-26}$ | 2.087 $10^{-31}$ |
| 600 | 1.966 $10^{-2}$ | 2.545 $10^{-1}$ | 2.576 $10^{-23}$ | 6.875 $10^{-24}$ | 8.753 $10^{-29}$ |
| 700 | 3.275 $10^{-2}$ | 5.650 $10^{+1}$ | 9.526 $10^{-21}$ | 2.276 $10^{-21}$ | 2.061 $10^{-25}$ |
| 800 | 4.720 $10^{-2}$ | 3.306 $10^{+3}$ | 8.033 $10^{-19}$ | 1.767 $10^{-19}$ | 6.968 $10^{-23}$ |
| 900 | 6.190 $10^{-2}$ | 7.934 $10^{+4}$ | 2.528 $10^{-17}$ | 5.215 $10^{-18}$ | 6.458 $10^{-21}$ |
| 1000 | 7.613 $10^{-2}$ | 1.017 $10^{+6}$ | 3.985 $10^{-16}$ | 7.820 $10^{-17}$ | 2.419 $10^{-19}$ |

Table 1. Tracer diffusion coefficient (m$^2$ s$^{-1}$) and correlation factor for Y in BCC iron. Exact results are compared with those obtained by Barouh [2] and with selfdiffusion.

The results of our calculation are displayed in Table 1. For this complex mechanism, the Arrhenius plot of the correlation factor $f_Y^{exact}$ has a slight downward curvature: the apparent activation energy increases from 0.16 eV at higher temperatures to 0.20 at the lower ones. The Arrhenius plot of the macrojump frequency $\Gamma_{MJ}^{SOB}$ exhibits the reverse trend: the apparent activation energy decreases from 1.98 eV at high temperatures down to 1.94 eV at the lower ones. The two effects cancel out each other in the product and the activation energy of the diffusion coefficient is found equal to 2.14 eV over the whole range of temperature with no noticeable curvature. The tracer diffusion coefficient of Y is definitely larger than the self-diffusion coefficient in BCC iron displayed in the last column.

A first approximate evaluation of yttrium diffusivity was recently proposed by Barouh [2]. The approximation consists in ignoring the 1nn↔5nn transitions ($W_3^{''}, W_4^{''}$) which leads to flicker events which produce no net transport of the yttrium atom, and the 1↔2 transitions ($W_3, W_4$ ) which require too high an energy. Only the 1nn↔3nn jump (frequency $W_3^{'}$) and the reverse jump (frequency $W_4^{'}$) are kept; at last, for lack of anything better, a constant correlation factor $f_Y^{Barouh} = 0.5$ was assumed. The diffusion coefficient $D_{Y*}^{Barouh}$ is then expressed by a single thermally activated term $D_O \exp\left(-E_{act} / (k_B T)\right)$, with $E_{act} = 2.10\ eV$ and $D_O = 3.0\ 10^{-6}\ \mathrm{m}^2 s^{-1}$. Surprisingly, this approximation is smaller only by a factor ranging from 1.7 at the lower temperatures to 5.1 at the higher ones, notwithstanding the constant value adopted for the correlation factor, at variance with the exact one. This unexpected agreement deserves some comment.

We show in Appendix E that Barouh's approximation induces automatically a (nearly) constant correlation factor. Using this approximation (denoted later on with the superscript 'SOB' meaning 'in the spirit of Barouh') in our exact calculation of $Q$ is made by preventing the vacancy to take the ignored paths: the migration energies for $W_3, W_4$ and $W_3^{''}, W_4^{''}$ are set equal to an arbitrary high value (say, 100.0 eV), in order to prevent the corresponding transitions to enter the numerical result in a detectable way. However, no further assumption is made on the value of the correlation factor $f_Y^{SOB}$ itself, which remains the result of the calculation. The expression of the approximated macrojump frequency $\Gamma_{MJ}^{SOB}$ can be easily extracted from Eq. (25):

$$\Gamma_{MJ}^{SOB} \approx \frac{8C_{V0}\left(3e^{\beta E_3} W_4^{'}\right)\left(6W_3^{'}\right)}{8C_{V0}\left(3e^{\beta E_3} W_4^{'}\right) + \left(6W_3^{'}\right)}.$$

The numerical evaluation shows that the term proportional to $C_{V0}$ in the denominator is always negligible, which leads to $\Gamma_{MJ}^{SOB} \approx 24 C_{V0} e^{\beta E_3} W_4^{'}$ and to an approximated diffusivity equal to $D_{Y*}^{SOB} = \Gamma_{MJ}^{SOB} f_Y^{SOB} \omega^2 / 12 = 2C_{V0}\, e^{\beta E_3} W_4^{'}\, f_Y^{SOB} \omega^2$.

Table 2 displays the values of $f_Y^{SOB}$ and $\Gamma_{MJ}^{SOB}$. Since $f_Y^{SOB}$ is practically temperature independent, the total effective activation energy for yttrium diffusion reduces to $E_F^V - E_3 + E_M^{4'}$, which is equal to 2.12 - 0.14 + 0.16 = 2.14 eV.

In the particular case under consideration, a further striking point is the fact that the product $f_Y^{SOB}\Gamma_{MJ}^{SOB}$ is very close to the exact one $f_Y^{exact}\Gamma_{MJ}^{exact}$ over the whole temperature range, with a tendency to depart progressively at higher temperatures: the noticeable underestimation of $\Gamma_{MJ}^{SOB}$ is totally compensated by the too large value of the correlation factor. This ii the root of the success of Barouh's approximation.

The conclusion of this section points out the fact that the yttrium atom is definitely more rapid than the iron atom in the bcc structure.

One additional comment is worth being made, about the smallness of the correlation factor $f_Y^{exact}$: in Appendix F, it is shown that this small value is not implied by the existence of the Y-vacancy complex and its role in the migration mechanism, but only by the set of vacancy jump frequencies around the yttrium atom. The OSA occupies alternately lattice sites and midpoints of the first neighbour bonds. Considering this mechanism as a new one, independently of any energetic consideration, and assuming that all jump frequencies are equal to a common value, i.e. $W_{i \to j}^{shell} = W_0 \quad \forall\, i, j$, then the correlation factor for this new mechanism becomes equal to 0.761603, that is, close to the value of the correlation factor for self-diffusion with a pure vacancy mechanism in the BCC structure.

| T(K) | $f_Y^{SOB}$ | $\Gamma_{MJ}^{SOB}$ | $D_{Y*}^{SOB}$ | $\frac{D_{Y*}^{SOB}}{D_{Y*}^{exact}}$ |
|---|---|---|---|---|
| 300 | 0.3333 | $1.587\ 10^{-20}$ | $2.723\ 10^{-41}$ | 1.000 |
| 400 | 0.3333 | $1.543\ 10^{-11}$ | $2.648\ 10^{-32}$ | 1.000 |
| 500 | 0.3333 | $3.812\ 10^{-6}$ | $6.542\ 10^{-27}$ | 1.000 |
| 600 | 0.3333 | $1.501\ 10^{-2}$ | $2.575\ 10^{-23}$ | 1.000 |
| 700 | 0.3333 | 5.550 | $9.525\ 10^{-21}$ | 1.000 |
| 800 | 0.3334 | $4.679\ 10^{+2}$ | $8.032\ 10^{-19}$ | 1.000 |
| 900 | 0.3337 | $1.471\ 10^{+4}$ | $2.527\ 10^{-17}$ | 1.000 |
| 1000 | 0.3341 | $2.314\ 10^{+5}$ | $3.980\ 10^{-16}$ | 0.999 |

Table 2. Comparison between the exact result and the SOB approximation for yttrium in bcc iron: values of effective macrojump frequency and correlation factor.

## III Diffusion in a FCC lattice

Only the differences with the case of the BCC lattice examined above are pointed out.

The rotation of the divacancy of Fig. 3 is obtained when the vacancy at the centre of the cell jumps toward the iron atom located in the forefront lattice plane. During the ascent toward the saddle configuration, the OSA is pushed back on the substitutional site; it then relaxes toward the midpoint of the new bond occupied by the divacancy at the end of the jump. Thus, the net displacement of the OSA takes place from an intermediate site to a neighbouring one, that is, without interrupting the macrojump in progress. Thanks to this elementary jump, the macrojump length can be arbitrary long and its length depends only on the relative values of the frequencies $W_{II}$ and $W_{IS}$. Fig. 5 displays the 8 intermediate sites which are first neighbours of the starting one.

### III-1 Calculating the average squared displacement during one encounter in the FCC lattice

The previous recurrence equations must be modified to take into account a new type of jump, namely the jump of frequency $W_{II}$ depicted on Fig. 3. The first equation is kept as is, since performing an S→I jump needs necessarily that the previous one was of I→S type, hence :

$SI_n^{\lambda_i}(r) = \sum_{\{\lambda_j\}} IS_{n-1}^{-\lambda_i}(r+\lambda_j)p_{-\lambda_j,\lambda_i}$ where $\{\omega_i\}$ stand now for the twelve jump vectors in the FCC lattice and $\{\lambda_i\} = \{\omega_i/2\}$.

The second equation takes into account the fact that the jump which precedes an I→S jump is no longer necessarily an S→I jump; it can also be an I→I one. The jump frequency of the B* atom for S→I, I→I and I→S jump are denoted $W_{SI}$, $W_{II}$, $W_{IS}$ respectively. Parameters $\alpha = W_{IS}/(2W_{IS}+8W_{II})$ and $\beta = W_{II}/(2W_{IS}+8W_{II})$ are defined with $2\alpha + 8\beta = 1$. Hence:

$$IS_n^{-\lambda_i}(r+\lambda_i) = \alpha\left(SI_{n-1}^{\lambda_i}(r) + SI_{n-1}^{-\lambda_i}(r+\omega_i) + \sum_{\{\lambda_{ic}\}} II_{n-1}^{-\lambda_{ic}}(r+\lambda_i+\lambda_{ic})\right) \qquad (27)$$

where $\{\lambda_{ic}\}$ are the vectors connecting the intermediate site $r+\lambda_i$ to the eight intermediate neighbours $r+\lambda_i+\lambda_{ic}$. The vectors $\{\lambda_{ic}\}$ are partionned into $\{\lambda_{ic}^+\}$ and $\{\lambda_{ic}^-\} = -\{\lambda_{ic}^+\}$ as depicted on Figure 5. They obey the conditions $\overrightarrow{\lambda_{ic}^+}.\overrightarrow{\lambda_i} = +\lambda^2/2$ and $\overrightarrow{\lambda_{ic}^-}.\overrightarrow{\lambda_i} = -\lambda^2/2$ and can be easily identified to vectors which belong to the set $\{\lambda_j\}$ and which obey $\overrightarrow{\lambda_j^+}.\overrightarrow{\lambda_i} = +\lambda^2/2$ and $\overrightarrow{\lambda_j^-}.\overrightarrow{\lambda_i} = -\lambda^2/2$. Hence $\{\lambda_{ic}^+\}$ and $\{\lambda_{ic}^-\}$ are renamed $\{\lambda_j^+\}$ and $\{\lambda_j^-\}$. As a result, the locations of the eight intermediate neighbours can be equally well defined by the vectors $\{r+\lambda_j^+\}$ and $\{r+2\lambda_i+\lambda_j^-\} = \{r+2\lambda_i-\lambda_j^+\}$. Hence the identity:

$$\sum_{\{\lambda_{ic}\}} II_{n-1}^{-\lambda_{ic}}(r+\lambda_i+\lambda_{ic}) = \sum_{\{\lambda_j^+\}} II_{n-1}^{\lambda_i-\lambda_j^+}(r+\lambda_j^+) + \sum_{\{\lambda_j^-\}} II_{n-1}^{-\lambda_i-\lambda_j^-}(r+2\lambda_i+\lambda_j^-) \qquad (28)$$

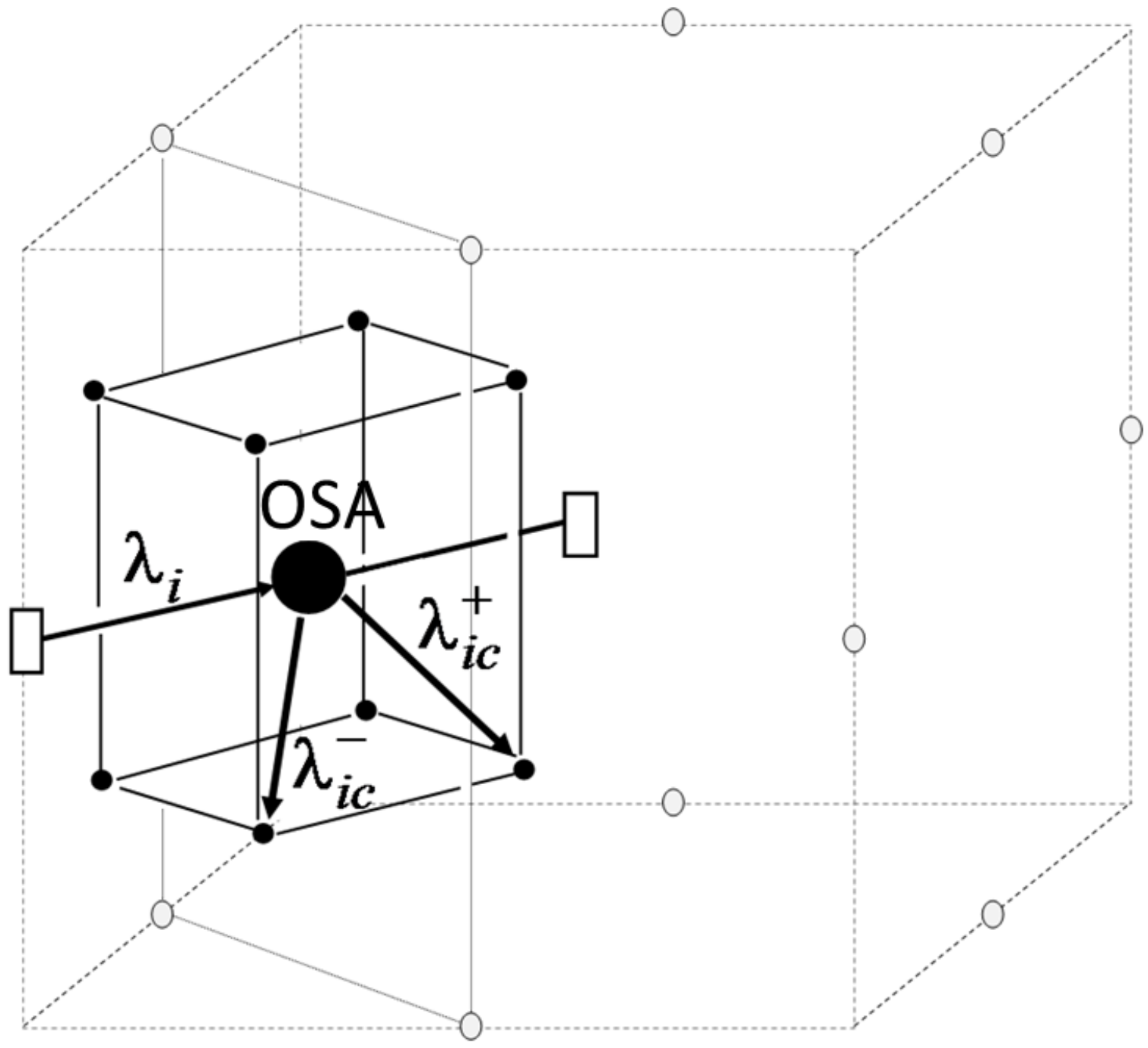

Figure 5. Jump vectors from an intermediate site to its eight neighbours in a FCC lattice

At last, the II functions can be eliminated at the benefit of IS functions. Indeed when the Y atom sits on an intermediate site $r+\lambda_j^+$ it has a probability $\alpha$ of choosing an I→S jump carrying it to lattice site $r$ and a probability $\beta$ of choosing the I→I jump carrying it at $(r+\lambda_i)$. Hence $II_{n-1}^{\lambda_i-\lambda_j^+}(r+\lambda_j^+)=\frac{\beta}{\alpha}IS_{n-1}^{-\lambda_j^+}(r+\lambda_j^+)$. With the same remark $II_{n-1}^{-\lambda_i-\lambda_j^-}(r+2\lambda_i+\lambda_j^-)$ is replaced by $\frac{\beta}{\alpha}IS_{n-1}^{-\lambda_j^-}(r+2\lambda_i+\lambda_j^-)$. The above recurrence equation becomes:

$$\begin{aligned}IS_n^{-\lambda_i}(r+\lambda_i)&=\alpha\, SI_{n-1}^{\lambda_i}(r)\ +\alpha\, SI_{n-1}^{-\lambda_i}(r+2\lambda_i)\\&+\beta\sum_{\{\lambda_j^+\}}IS_{n-1}^{-\lambda_j^+}(r+\lambda_j^+)\ +\beta\sum_{\{\lambda_j^-\}}IS_{n-1}^{-\lambda_j^-}(r+2\lambda_i+\lambda_j^-)\cdot\end{aligned}\tag{29}$$

The moments are evaluated in Appendix G with the final result:

$$\begin{aligned}<R^2>_{Enc}&=\frac{12M_{IS2}\,\lambda^2}{12M_{IS0}}=\frac{1+2\alpha+4\alpha Q}{2\alpha(1-P)}\lambda^2=\frac{1+2\alpha+4\alpha Q}{8\alpha(1-P)}\omega^2\\&=\frac{(1+2\alpha)}{8\alpha(1-P)}(1+\frac{4\alpha}{1+2\alpha}Q)\omega^2\end{aligned}\tag{30}$$

As above, the average square length of the macrojump is found equal to:

$$<R^2>_{MJ}=\frac{1+2\alpha}{2\alpha}\lambda^2,\tag{31}$$

the random square displacement made of $(1-P)^{-1}$ macrojump equals :

$$<R^2>_{Rand}=\frac{1}{1-P}\frac{1+2\alpha}{2\alpha}\lambda^2,\tag{32}$$

and the correlation factor is defined as usual by

$$f_B = \frac{< R^2 >_{Enc}}{< R^2 >_{Rand}} = 1 + \frac{4\alpha}{1+2\alpha} Q \qquad (33)$$

The smaller $\alpha$, the larger the average squared length of a macro-jump, the larger the displacement during an encounter and the larger the correlation factor. The latter becomes close to unity if the Y atom migrates essentially from an intermediate site to another without coming back on a lattice site, which mimics in a close way a direct interstitial mechanism.

**III-2 Calculating the time delay allotted to an encounter in the FCC lattice**

The average frequency $\Gamma_{SI}$ of an S→I jump is the total jump frequency of a vacancy towards a first neighbour site of the tracer. The sites $R_{init}$ it starts from belong to more distant shells labelled 'j' than the first one and the vacancy jump frequencies from shell 'j' to shell '1' are named $W_{j\to1}^{shell}$ (j= 2, 3, 4 for a FCC lattice). The probability of finding a vacancy on shell 'j' is by definition its atomic concentration $C_{Vj}$. The number $nbond_{1\to j}$ of bonds connecting one given site of the 1st shell to sites of the jth shell is given by $nbond_{1\to2} = 2,\ nbond_{1\to3} = 4,\ nbond_{1\to4} = 1$ and the frequency $\Gamma_{SI}$ is expressed as:

$$\Gamma_{SI} = zC_{V0} \sum_{jV1} nbond_{1\to j} e^{-\beta E_j} W_{j\to1}^{shell} \qquad (34)$$

The frequency $\Gamma_{IS}$ of the return onto a lattice site must take into account the I→I jumps with the frequency $W_{II}$ which postpone the final return through an S→I jump of frequency $W_{IS}$. The average residence time on an intermediate site is $\tau = \frac{1}{2W_{IS} + 8W_{II}}$. The time delay for the return on a lattice site is nothing but the total time spent on intermediate sites, that is, $\tau$ with probability $2\alpha$, $2\tau$ with probability $2\alpha(8\beta)$,… $n\,\tau$ with probability $2\alpha(8\beta)^{n-1}$…etc, which gives and average time delay equal to $\frac{2\alpha\tau}{(1-8\beta)^2} = \frac{\tau}{2\alpha} = \frac{1}{2W_{IS}}$. Thus, introducing I→I jumps enhances the number of visited intermediate sites from 1 to $\frac{1}{2\alpha}$; but the residence time on intermediate sites is decreased from $\frac{1}{2W_{IS}}$ to $\frac{1}{2W_{IS} + 8W_{II}} = \frac{1}{2W_{IS}} 2\alpha$, i.e. by the same factor. As a result the two compensate each other and, as above,

$$\Gamma_{IS} = 2W_{IS}\,. \qquad (35)$$

The final expression of the diffusion coefficient is then:

$$D_B^* = \frac{< R^2 >_{Enc}}{6\Delta t_{Enc}} = \frac{\Gamma_{SI}\Gamma_{IS}}{\Gamma_{SI} + \Gamma_{IS}} \frac{1+2\alpha+4\alpha Q}{48\alpha} \omega^2. \qquad (36)$$

**III-3 Calculating Q in the FCC lattice**

The proportions of the vacancy which are rejected on neighbouring sites belonging to 2nd, 3rd and 4th neighbour shells after an I→S jump are denoted:

$$c_3 = \frac{W^{shell}_{1\to2}}{W_{IS}} \qquad c_3^{'} = \frac{W^{shell}_{1\to3}}{W_{IS}} \qquad c_3^{''} = \frac{W^{shell}_{1\to4}}{W_{IS}} \tag{37}$$

with $W_{IS} = 2W^{shell}_{1\to2} + 4W^{shell}_{1\to3} + W^{shell}_{1\to4}$ and $2c_3 + 4c_3^{'} + c_3^{''} = 1$.

An initial condition complying with the four-fold symmetry around <100> is introduced. The vacancy separating from the B* atom through an outward jump from site $r = \omega_{110}$ is replaced by four quarters of vacancy starting from $r = \omega_{110}, \omega_{101}, \omega_{1\bar{1}0}, \omega_{10\bar{1}}$. A last a negative source is introduced along the negative x-axis on sites $r = \omega_{\bar{1}10}, \omega_{\bar{1}01}, \omega_{\bar{1}\bar{1}0}, \omega_{\bar{1}0\bar{1}}$ by mirror antisymmetry. The initial condition of the diffusion problem is given by:

$$\begin{aligned}
CI(r,0) = {} & \mathit{4}\ \frac{c_3}{4}\delta(r-\omega_{200}) - \mathit{4}\ \frac{c_3}{4}\delta(r-\omega_{\bar{2}00}) \\
& + \mathit{1}\ \frac{c_3^{'}}{4}\begin{pmatrix} \delta(r-\omega_{121}) + \delta(r-\omega_{12\bar{1}}) + \delta(r-\omega_{112}) + \delta(r-\omega_{11\bar{2}}) \\ +\delta(r-\omega_{1\bar{1}2}) + \delta(r-\omega_{1\bar{1}\bar{2}}) + \delta(r-\omega_{1\bar{2}1}) + \delta(r-\omega_{1\bar{2}\bar{1}}) \end{pmatrix} \\
& - \mathit{1}\ \frac{c_3^{'}}{4}\begin{pmatrix} \delta(r-\omega_{\bar{1}21}) + \delta(r-\omega_{\bar{1}2\bar{1}}) + \delta(r-\omega_{\bar{1}12}) + \delta(r-\omega_{\bar{1}1\bar{2}}) \\ +\delta(r-\omega_{\bar{1}\bar{1}2}) + \delta(r-\omega_{\bar{1}\bar{1}\bar{2}}) + \delta(r-\omega_{\bar{1}\bar{2}1}) + \delta(r-\omega_{\bar{1}\bar{2}\bar{1}}) \end{pmatrix} \\
& + \mathit{2}\frac{c_3^{'}}{4}\left(\delta(r-\omega_{211}) + \delta(r-\omega_{21\bar{1}}) + \delta(r-\omega_{2\bar{1}1}) + \delta(r-\omega_{2\bar{1}\bar{1}})\right) \\
& - \mathit{2}\frac{c_3^{'}}{4}\left(\delta(r-\omega_{\bar{2}11}) + \delta(r-\omega_{\bar{2}1\bar{1}}) + \delta(r-\omega_{\bar{2}\bar{1}1}) + \delta(r-\omega_{\bar{2}\bar{1}\bar{1}})\right) \\
& + \mathit{1}\ \frac{c_3^{''}}{4}\left(\delta(r-\omega_{220}) + \delta(r-\omega_{202}) + \delta(r-\omega_{2\bar{2}0}) + \delta(r-\omega_{20\bar{2}})\right) \\
& - \mathit{1}\ \frac{c_3^{''}}{4}\left(\delta(r-\omega_{\bar{2}20}) + \delta(r-\omega_{\bar{2}02}) + \delta(r-\omega_{\bar{2}\bar{2}0}) + \delta(r-\omega_{\bar{2}0\bar{2}})\right)
\end{aligned}$$

where the italic digit reflects the number of those sites belonging to the first neighbour shell which contribute to the presence of the vacancy on a given site of the 2nd, 3rd or 4th neighbour shell. The contributions arriving on plane x=0 cancel out each other thanks to the mirror plane and disappear from the formula. Using the numbering of subsets corresponding to a solute-vacancy interaction up to 7th neighbour shell, the Fourier transform of initial condition becomes:

$$FCI(k,0) = \frac{c_3}{4} nlink_{10\to9}\, f_{10} + \frac{c_3^{'}}{4} nlink_{11\to9}\, f_{11} + \frac{c_3^{'}}{4} nlink_{12\to9}\, f_{12} + \frac{c_3^{''}}{4} nlink_{13\to9}\, f_{13} \tag{38}$$

The value of the average cosine will then be equal to:

$$\begin{aligned}
Q^{FCC} = {} & -n^{+}_{10} LL^{+}_{10}(nlink_{10\to9} W^{sub}_{10\to9}) - n^{+}_{11} LL^{+}_{11}(nlink_{11\to9} W^{sub}_{11\to9}) \\
& - n^{+}_{12} LL^{+}_{12}(nlink_{12\to9} W^{sub}_{12\to9}) - n^{+}_{13} LL^{+}_{13}(nlink_{13\to9} W^{sub}_{13\to9})
\end{aligned} \tag{39}$$

In the FCC lattice $nlink_{10\to9} = 4$ , $nlink_{11\to9} = 1$ , $nlink_{12\to9} = 2$ , $nlink_{13\to9} = 1$, $n^{+}_{10} = 1$ , $n^{+}_{11} = 8$ , $n^{+}_{12} = 4$ , $n^{+}_{13} = 4$, and according to the standard notations $W^{sub}_{10\to9} = W^{shell}_{2\to1} = W_4$ , $W^{sub}_{11\to9} = W^{sub}_{12\to9} = W^{shell}_{3\to1} = W_4^{'}$, $W^{sub}_{13\to9} = W^{shell}_{4\to1} = W_4^{''}$ this yields the final expressions :

$$FCI(k,0) = c_3 f_{10} \qquad + \frac{c_3^{'}}{4} f_{11} \qquad + \frac{c_3^{'}}{2} f_{12} \qquad + \frac{c_3^{"}}{4} f_{13},$$

$$Q^{FCC} = -4LL_{10}^{+} W_4 - 8(LL_{11}^{+} + LL_{12}^{+}) W_4^{'} - 4LL_{13}^{+} W_4^{"}.$$

The average frequencies entering the macro-jump frequency are given by:

$$\Gamma_{SI} = 12C_{V0}\left(2e^{-\beta E_2}W_4 + 4e^{-\beta E_3}W_4^{'} + e^{-\beta E_4}W_4^{"}\right), \qquad (40)$$

$$\Gamma_{IS} = 2\left(2W_3 + 4W_3^{'} + W_3^{"}\right). \qquad (41)$$

The final explicit expression of the tracer diffusion coefficient is:

$$D_B^{*} = \frac{C_{V0}\left(2e^{-\beta E_2}W_4 + 4e^{-\beta E_3}W_4^{'} + e^{-\beta E_4}W_4^{"}\right)\left(2W_3 + 4W_3^{'} + W_3^{"} + 2W_{II}\right)}{6C_{V0}\left(2e^{-\beta E_2}W_4 + 4e^{-\beta E_3}W_4^{'} + e^{-\beta E_4}W_4^{"}\right) + \left(2W_3 + 4W_3^{'} + W_3^{"}\right)} f_B\, \omega^2. \qquad (42)$$

## III-4 Application to the case of Y in FCC iron

In this section, iron is assumed to be non-magnetic. The lattice parameter is set equal to 3.51 $10^{-10}$ m. The vacancy formation and migration energies are found equal to 2.543 eV and 1.34 eV respectively. The vacancy formation entropy is arbitrarily taken equal to $2k_B$, which is a commonly accepted value in FCC metals [11]. The interaction energies and migration barriers are calculated in the same way as for the BCC lattice. The pre-exponential term is taken equal to the Debye frequency $10^{13}$ $s^{-1}$ for all jump frequencies. All the details on the first-principles calculations will be reported in a forthcoming publication [12].

As above, the large attractive interaction energy between the Y atom and the vacancy is accompanied by small dissociation frequencies and high re-association ones. All interaction energies are displayed in Fig. 6; the migration energies are obtained by subtracting the energy of the starting configurations from the energy of the saddle configurations. The additional feature is the existence of the rotation frequency $W_{II} = W_1$ : the corresponding migration energy is equal to the saddle configuration energy (1.15 eV) displayed in Fig. 6 minus the interaction energy at first neighbor distance (-1.32 eV), i.e. 2.47 eV. Table B2 shows that the neighbours of a site on the 7$^{th}$ shell belong to shells 3 to 13: not all of these neighbours are displayed for sake of simplicity, since the vacancy migration energy has been set equal to the bulk value 1.34 eV for all the transitions $W_{7\to j}^{shell}$ with $j \geq 7$.

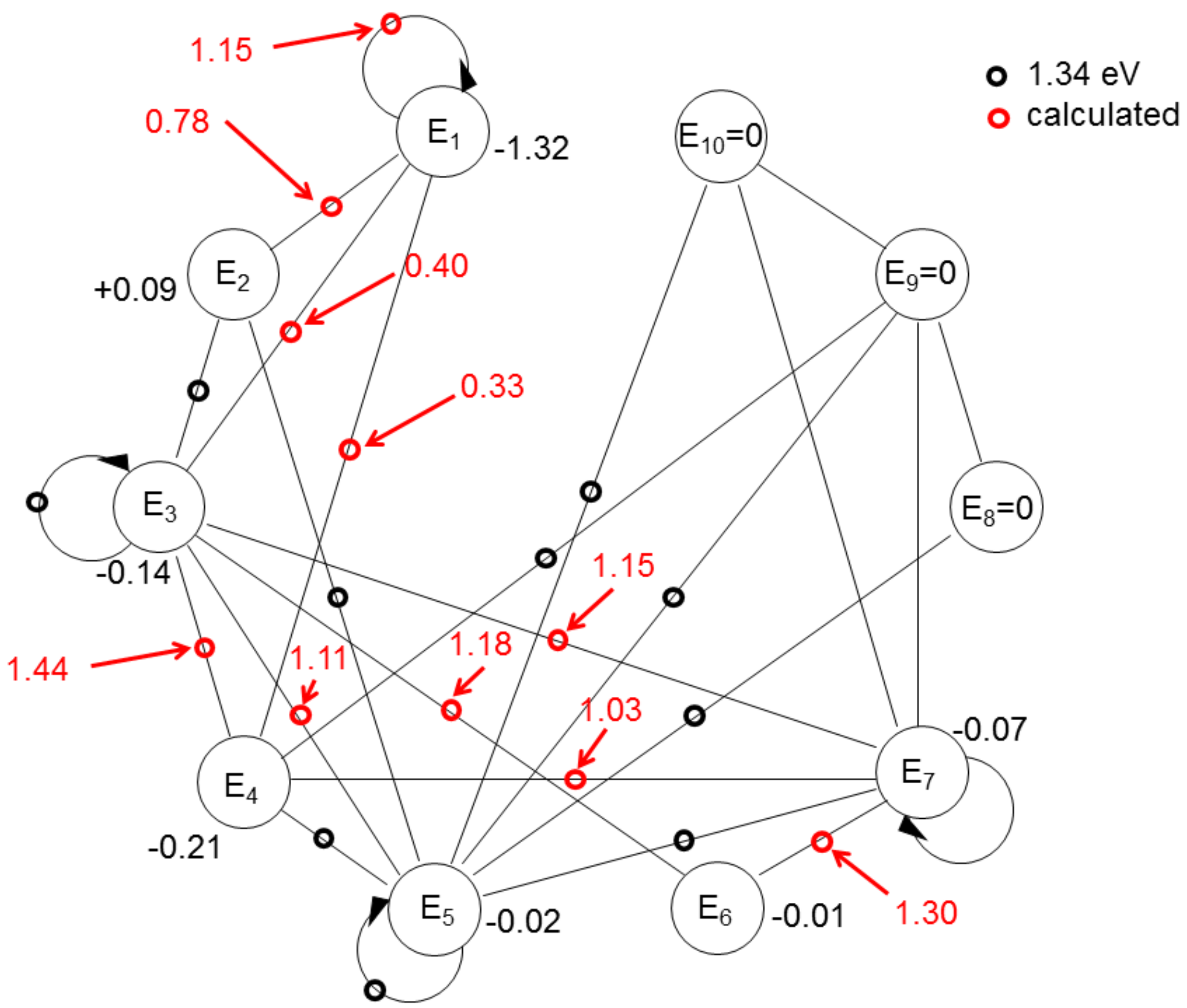

Figure 6. Diagram of interaction and saddle configuration energies for Y in FCC iron (eV). Data in black are the Y-V interactions at rest; data in red are the calculated saddle energies for the migration from $E_i$ to $E_j$ ; black saddle positions are set equal to the vacancy migration barrier in the bulk.

In Table 3 below are gathered the values of the correlation factor $f_Y^{exact}$ and diffusion coefficient $D_{Y*}^{exact}$, to be compared with the values of self-diffusion in FCC iron. Correlation effects are noticeable, but less marked than in the BCC structure.
In the present case, the role of the rotation frequency $W_{II} = W_1$ is negligible, because of its large activation barrier: the factor $\alpha$ remains very close to 1/2.

As a conclusion, Y atom diffuses more rapidly than Fe in the FCC phase.

| T(K) | $f_Y^{exact}$ | $D_{Y*}^{exact}$ | $D_{Fe*}$ |
|---|---|---|---|
| 900 | 0.1555 | $2.528\ 10^{-23}$ | $1.733\ 10^{-28}$ |
| 1000 | 0.1614 | $1.130\ 10^{-21}$ | $2.591\ 10^{-26}$ |
| 1100 | 0.1664 | $2.538\ 10^{-20}$ | $1.558\ 10^{-24}$ |
| 1200 | 0.1710 | $3.405\ 10^{-19}$ | $4.733\ 10^{-23}$ |
| 1300 | 0.1754 | $3.074\ 10^{-18}$ | $8.503\ 10^{-22}$ |
| 1400 | 0.1797 | $2.033\ 10^{-17}$ | $1.011\ 10^{-20}$ |
| 1500 | 0.1841 | $1.049\ 10^{-16}$ | $8.645\ 10^{-20}$ |
| 1600 | 0.1888 | $4.428\ 10^{-16}$ | $5.652\ 10^{-19}$ |

Table 3. Tracer diffusion coefficient ($m^2\ s^{-1}$) and correlation factor for Y in FCC iron. The results are compared with the self-diffusion data.

## IV Conclusions

Thanks to a new theoretical analysis of OSA diffusion, the following points were established:

*OSA can form tightly bound complexes with a vacancy: the OSA sits on an intermediate site between the two adjacent vacancies which are at the ends of a first neighbor bond;

*the first stage of the OSA displacement takes place during the formation of the complex, when the OSA is pushed from a substitutional onto an intermediate location (SI displacement); in the BCC lattice, the next stage is necessarily a dissociation of the complex which brings back the OSA on a substitutional site (IS displacement); this displacement cancels or consolidates the previous SI displacement; in the FCC lattice, the next stage can also be a rotation of the complex, which displaces the OSA from on intermediate site to a neighboring one (II displacement). The total displacement of the OSA from its starting substitutional site to the arrival substitutional one is called a macrojump, which becomes the new elementary displacement;

*all the types of jump (SI, II, IS) are associated with corresponding displacement functions, which are closely interlinked by recurrence relationships. The second order moments of these functions yield the quadratic displacement of the OSA as a function of the number of macrojumps, the average quadratic length of a macrojump, together with the time required to perform it. This gives the final expression of the diffusion coefficient.

*there is no correlation between the displacements belonging to the same macrojump; the only correlation effect arises between the IS displacement of a macrojump and the SI displacement of the next one; it depends of the return probabilities of the vacancy in the neighborhood of the OSA;

*these return probabilities are evaluated through a Laplace and Fourier transform of a master equation describing the vacancy displacement around the OSA; these probabilities are shown to be the solution of a linear system, the size of which is determined by the range of the OSA-vacancy interaction;

*when applied to the case of yttrium in iron in the BCC and FCC structure, thanks to the calculation of all migration barriers by first-principles methods, the present approach shows that Y is a very rapid diffuser; the correlation effects are large (the correlation factor is small), but they are overcompensated by a very high macrojump frequency, yielding finally a diffusivity for yttrium which is always orders of magnitude larger than that for iron in the BCC and FCC structures.

*most probably this conclusion can be extended: all OSAs forming a complex with the vacancy in cubic lattices are rapid diffusers.

## Acknowledgements

One of us (JLB) thanks the LRC-Méso (a joint laboratory formed by CMLA and CEA/DAM) for his support during the study.

## Bibliography

[1] A. Claisse, Olsson, 2013, Nuclear Instruments and Methods in Physics Research B, 303, p. 18.
[2] C. Barouh, 2015, ‘Modélisation multi-échelle de l’interaction entre les éléments d’alliage et les lacunes dans les aciers ferritiques’, PhD, Université d’Orléans, France.
[3] D.J. Hepburn, E. MacLeod and G.J. Ackland, 2015, Phys. Rev. B92, p. 014110.
[4] H. Höhler, N. Atodiresei, K. Schroeder, R. Zeller and P.H. Dederichs, 2004, Phys. Rev. B70, p. 155313.
[5] A.D. Le Claire, 1970, Physical Chemistry, An Advanced Treatise, **Vol. 10**, H. Eyring, W. Anderson and W. Jost Ed. (Academic Press, NY 1970), p. 261.
[6] J.L. Bocquet, 2014, Correlation factor for diffusion in cubic crystals with solute–vacancy interactions of arbitrary range (Appendix A) in arxiv:1312.4390
[7] J.L. Bocquet, 2014, Philos. Mag., 94, p. 3603 and http://dx.doi.org/10.1080/14786435.2014.965768
[8] E.W. Montroll, 1956, J. Soc. Indus. App. Math. 4, p. 241.
[9] D. Wolf, 1983, Phil. Mag. A47, p. 147.
[10] P. Benoist, J.L. Bocquet, P. Lafore, 1977, Acta Metallurgica 25, p. 265.
[11] J.J. Burton, 1972, Phys. Rev. B5, p. 2948.
[12] Chu Chun Fu, 2016, private communication; to be published.

# APPENDIX A
# Mean square displacement during an encounter

Zeroth order moments for SI and IS functions :

$$\sum_{\{r\}}\sum_{n=2}^{\infty} SI_n^{\lambda_i}(r) = \sum_{\{\lambda_j\}}\sum_{\{r\}}\sum_{n=2}^{\infty} IS_{n-1}^{-\lambda_j}(r+\lambda_j)\,p_{-\lambda_j,\lambda_i}$$

$$M_{SI0} - \frac{1}{z} = \sum_{\{\lambda_j\}} p_{-\lambda_j,\lambda_i}\sum_{\{r\}}\sum_{n=2}^{\infty} IS_{n-1}^{-\lambda_j}(r+\lambda_j) = \sum_{\{\lambda_j\}} p_{-\lambda_j,\lambda_i} M_{IS0}$$

Hence : $M_{SI0} - \frac{1}{z} = PM_{IS0}$ (A1)

where $P = \sum_{\{\lambda_j\}} p_{-\lambda_j,\lambda_i}$ (A2)

is the total probability that the solute atom B*, after its return on a substitutionnal lattice site, will be later on pushed again onto an intermediate site by the same vacancy.

$$\sum_{\{r\}}\sum_{n=2}^{\infty} IS_n^{-\lambda_i}(r+\lambda_i) = \frac{1}{2}\sum_{\{r\}}\sum_{n=2}^{\infty}\left(SI_{n-1}^{\lambda_i}(r) + SI_{n-1}^{-\lambda_i}(r+\omega_i)\right)$$

$$M_{IS0} = \frac{1}{2}\left(M_{SI0} + M_{SI0}\right)$$

hence : $M_{SI0} = M_{IS0} = \dfrac{1}{z(1-P)}$ . (A3)

The average number of S→I jumps performed by the tracer atom in a particular direction during its encounter with a given vacancy is equal to the number of I→S jumps along the same direction. The multiplicative factor 1/z reminds that the moment is calculated for a function IS or SI with a well defined orientation. The average total number of jumps performed by the B* atom (in all possible directions) is thus equal to $zM_{SI0} = \dfrac{1}{1-P} = 1 + P + P^2 ...$

First order moments for SI and IS functions:
they are vectors aligned with the superscripts of the two functions IS and SI :

$$\sum_{\{r\}}\sum_{n=2}^{\infty}(r+\lambda_i) SI_n^{\lambda_i}(r) = \sum_{\{\lambda_j\}}\sum_{\{r\}}\sum_{n=2}^{\infty}(r+\lambda_i) IS_{n-1}^{-\lambda_j}(r+\lambda_j)\,p_{-\lambda_j,\lambda_i}$$

$$M_{SI1}\overrightarrow{\lambda_i} - \frac{\overrightarrow{\lambda_i}}{z} = \sum_{\{\lambda_j\}} p_{-\lambda_j,\lambda_i} M_{IS1}(-\overrightarrow{\lambda_j}) + PM_{IS0}\overrightarrow{\lambda_i}$$

After a scalar product of the last line with $\overrightarrow{\lambda_i}$ :

$$\left(M_{SI1}\overrightarrow{\lambda_i} - \frac{\overrightarrow{\lambda_i}}{z}\right)\cdot\overrightarrow{\lambda_i} = M_{IS1}\sum_{\{\lambda_j\}} p_{-\lambda_j,\lambda_i}(-\overrightarrow{\lambda_j}.\overrightarrow{\lambda_i}) + PM_{IS0}\overrightarrow{\lambda_i}.\overrightarrow{\lambda_i}$$

$$(M_{SI1} - \frac{1}{z})\lambda^2 = M_{IS1}\lambda^2 Q + PM_{IS0}\lambda^2$$

Hence $M_{SI1} - \frac{1}{z} = M_{IS1}Q + PM_{IS0}$ , (A4)

with $Q=\sum_{\{\lambda_j\}} p_{-\lambda_j,\lambda_i}\left(\frac{-\overrightarrow{\lambda_j}.\overrightarrow{\lambda_i}}{\lambda^2}\right)=\sum_{\{\lambda_j\}} p_{-\lambda_j,\lambda_i}\cos(-\overrightarrow{\lambda_j},\overrightarrow{\lambda_i})$. (A5)

$Q$ is a weighted average cosine between two successive jumps. In the present case the restriction consists in considering only jump pairs (I→S, S→I). The pair (S→I, I→S) does not imply any correlation effect since the tracer on an intermediate site I has an equal probability of jumping towards the two ends of the bond.

$$\sum_{\{r\}}\sum_{n=2}^{\infty} r\, IS_n^{-\lambda_i}(r+\lambda_i)=\frac{1}{2}\sum_{\{r\}}\sum_{n=2}^{\infty} r\left(SI_{n-1}^{\lambda_i}(r)+SI_{n-1}^{-\lambda_i}(r+\omega_i)\right)$$

$$=\frac{1}{2}\sum_{\{r\}}\sum_{n=2}^{\infty}\begin{pmatrix}(r+\lambda_i)SI_{n-1}^{\lambda_i}(r)-\lambda_i SI_{n-1}^{\lambda_i}(r)\\ +(r+\omega_i-\lambda_i)SI_{n-1}^{-\lambda_i}(r+\omega_i)-(\omega_i-\lambda_i)SI_{n-1}^{-\lambda_i}(r+\omega_i)\end{pmatrix}$$

Introducing the notation for first order moments leads to

$$-M_{IS1}\overrightarrow{\lambda_i}=\frac{1}{2}\begin{pmatrix}M_{SI1}\overrightarrow{\lambda_i}-M_{SI0}\overrightarrow{\lambda_i}\\ -M_{SI1}\overrightarrow{\lambda_i}-M_{SI0}(\overrightarrow{\omega_i}-\overrightarrow{\lambda_i})\end{pmatrix}=\frac{1}{2}\left(-M_{SI0}\overrightarrow{\lambda_i}-M_{SI0}(\overrightarrow{\omega_i}-\overrightarrow{\lambda_i})\right)=-M_{SI0}\overrightarrow{\lambda_i}$$

or $M_{IS1}=M_{SI0}$. (A6)

Plugging this relation into the previous one gives :

$$M_{SI1}-\frac{1}{z}=M_{SI0}Q+PM_{IS0}=M_{IS0}(P+Q)$$

or $M_{SI1}=\dfrac{1+Q}{z(1-P)}$. (A7)

Second order moments for SI functions :

$$\sum_{\{r\}}\sum_{n=2}^{\infty}(r+\lambda_i)^2 SI_n^{\lambda_i}(r)=\sum_{\{\lambda_j\}} p_{-\lambda_j,\lambda_i}\sum_{\{r\}}\sum_{n=2}^{\infty}(r+\lambda_i)^2 IS_{n-1}^{-\lambda_j}(r+\lambda_j)$$

$$=\sum_{\{\lambda_j\}} p_{-\lambda_j,\lambda_i}\sum_{\{r\}}\sum_{n=2}^{\infty}(r^2+2r.\lambda_i+\lambda^2)IS_{n-1}^{-\lambda_j}(r+\lambda_j)$$

Introducing the notations of second order moments gives :

$$(M_{SI2}-\frac{1}{z})\lambda^2=\sum_{\{\lambda_j\}} p_{-\lambda_j,\lambda_i}\left\{M_{IS2}\lambda^2-2\lambda_i.M_{IS1}\lambda_j+M_{IS0}\lambda^2\right\}$$

or $M_{SI2}-\dfrac{1}{z}=M_{IS2}P+2M_{IS1}Q+M_{IS0}P$

For IS functions :

$$\sum_{\{r\}}\sum_{n=2}^{\infty} r^2 IS_n^{-\lambda_i}(r+\lambda_i)=$$

$$\frac{1}{2}\sum_{\{r\}}\sum_{n=2}^{\infty}\begin{pmatrix}\left\{(r+\lambda_i)^2-2\lambda_i.(r+\lambda_i)+\lambda^2\right\}SI_{n-1}^{\lambda_i}(r)\\ +\left\{(r+\omega_i-\lambda_i)^2-2(\omega_i-\lambda_i).(r+\omega_i-\lambda_i)+(\omega_i-\lambda_i)^2\right\}SI_{n-1}^{-\lambda_i}(r+\omega_i)\end{pmatrix}$$

Then

$$M_{IS2}\lambda^2 = \begin{pmatrix} \frac{1}{2}M_{IS2}\lambda^2 - \lambda_i.M_{SI1}\lambda_i + \frac{1}{2}M_{SI0}\lambda^2 \\ \frac{1}{2}M_{IS2}\lambda^2 - \lambda_i.(-M_{SI1}\lambda_i) + \frac{1}{2}M_{SI0}\lambda^2 \end{pmatrix}$$

$$= M_{IS2}\lambda^2 + M_{SI0}\lambda^2$$

or $M_{IS2} = M_{IS2} + M_{SI0}$ . (A8)

Hence :

$$M_{IS2}(1-P) = \frac{1}{z} + 2M_{IS0}Q + M_{IS0}(1+P) = \frac{2(1+Q)}{z(1-P)} \; . \tag{A9}$$

## Appendix B

## Neighbourhood relationships between subsets for a solute-vacancy interaction ranging up to 5th and 7th neighbour for bcc and fcc lattice respectively.

In the Tables B1-B2 below, subsets $[i^+]$ are listed in the 1rst column; the coordinates $(i_1, i_2, i_3)$ of the representative site are displayed in the 2nd column (with $0 \le i_2 \le i_3$); the neighbour shell it belongs to is given in 3rd column. The number of sites in subset $[i^+]$ is given in 4th column.

The subsets $[j]$ which are first neighbours of subset $[i^+]$ are listed in 5th column (numbers in square brackets) in ascending order. The number of bonds $nlink_{i \to j}$ connecting one given site of subset $[i^+]$ to sites of subset $[j]$ are displayed as a multiplicative factor followed by an 'x'.

A negative value displayed as $[-j]$ corresponds to a subset with negative x-coordinates and which has a mirror symmetry with subset $[j]$.

The symbol '∞' indicates that the targeted neighbour site is beyond the definition range of subsets and belongs to the background medium which is fully taken into account in the transport equation by the general term: the farther from the origin, the larger the number of targeted sites beyond the range of definition for the subsets.

Subset [0] contains only the origin of coordinates.

For the bcc lattice with interactions up to the 5th neighbour shell:

- subsets [1] to [4] correspond to subsets contained in the mirror-symmetry plane x=0.
- the last subset, the neighbours of which are entirely included in the present description, is subset number [11] belonging to 6th shell.

For the fcc lattice with interactions up to the 7th neighbour shell:

- subsets [1] to [8] correspond to subsets contained in the mirror-symmetry plane x=0.
- the last subset, the neighbours of which are entirely included in the present description, is subset number [20] belonging to 8th shell.

| $[i^+]$ | $(i_1,i_2,i_3)$ | Shell | $n_i^+$ | Neighbouring subsets [j] with multiplicity $nlink_{i\to j}$ | |
|---|---|---|---|---|---|
| [0] | 0,0,0 | 0 | 1 | 4x[-5] + 4x[5] | |
| [1] | 0,0,2 | 2 | 4 | 2x[-8] + 2x[-5] + 2x[5] + 2x[8] | |
| [2] | 0,2,2 | 3 | 4 | 1x[-12] + 2x[-8] +1x[-5] + 1x[5] + 2x[8] +1x[12] | |
| [3] | 0,0,4 | 6 | 4 | 2x[-18] + 2x[-8] + 2x[8] + 2x[18] | |
| [4] | 0,2,4 | 8 | 8 | 1x[-18] + 1x[-12] + 1x[-8] + 1x[8] + 1x[12] + 1x[18] | *+ 2x[∞]* |
| | | | | | |
| [5] | 1,1,1 | 1 | 4 | 1x[0] + 2x[1] + 1x[2] +1x[6] + 2x[7] + 1x[10] | |
| [6] | 2,0,0 | 2 | 1 | 4x[5] + 4x[9] | |
| [7] | 2,0,2 | 3 | 4 | 2x[5] + 2x[8] + 2x[9] + 2x[13] | |
| [8] | 1,1,3 | 4 | 8 | 1x[1] + 1x[2] + 1x[3] + 1x[4] + 1x[7] + 1x[10] +1x[14] + 1x[16] | |
| [9] | 3,1,1 | 4 | 4 | 1x[6] + 2x[7] + 1x[10] + 1x[11] + 2x[15] + 1x[17] | |
| [10] | 2,2,2 | 5 | 4 | 1x[5] + 2x[8] + 1x[9] +1x[12] + 2x[13] + 1x[19] | |
| [11] | 4,0,0 | 6 | 1 | 4x[9] + 4x[20] | |
| [12] | 1,3,3 | 7 | 4 | 1x[2] + 2x[4] + 1x[10] + 2x[16] | *+ 2x[∞]* |
| [13] | 3,1,3 | 7 | 8 | 1x[7] + 1x[10] + 1x[14] + 1x[15] + 1x[16] + 1x[17] | *+ 2x[∞]* |
| [14] | 2,0,4 | 8 | 4 | 2x[8] + 2x[13] + 2x[18] | *+ 2x[∞]* |
| [15] | 4,0,2 | 8 | 4 | 2x[9] + 2x[13] + 2x[20] | *+ 2x[∞]* |
| [16] | 2,2,4 | 9 | 8 | 1x[8] + 1x[12] + 1x[13] + 1x[18] + 1x[19] | *+ 3x[∞]* |
| [17] | 4,2,2 | 9 | 4 | 1x[9] + 2x[13] + 1x[19] + 1x[20] | *+ 3x[∞]* |
| [18] | 1,1,5 | 10 | 8 | 1x[3] + 1x[4] + 1x[14] + 1x[16] | *+ 4x[∞]* |
| [19] | 3,3,3 | 10 | 4 | 1x[10] + 2x[16] + 1x[17] | *+ 4x[∞]* |
| [20] | 5,1,1 | 10 | 4 | 1x[11] + 2x[15] + 1x[17] | *+ 4x[∞]* |

Table B1: The neighbour relationships between subsets in a BCC lattice with a solute-vacancy interaction range up to the 5$^{th}$ neighbour shell.

| $[i^+]$ | $(i_1,i_2,i_3)$ | Shell | $n_i^+$ | Neighbouring subsets [j] with multiplicity $nlink_{i\to j}$ |
|---|---|---|---|---|
| [0] | 0,0,0 | 0 | 1 | 4x[-9] + 4x[1] + 4x[9] |
| [1] | 0,1,1 | 1 | 4 | 2x[-11] +2x[-9] + 1x[0] + 2x[2] + 1x[3] + 2x[9] + 2x[11] |
| [2] | 0,0,2 | 2 | 4 | 1x[-14] + 2x[-11]+1x[-9] + 2x[1] + 2x[4] + 1x[9] + 1x[11] + 1x[14] |
| [3] | 0,2,2 | 4 | 4 | 2x[-17] + 2x[-11] + 1x[1] + 2x[4] + 1x[6] + 2x[11] + 2x[17] |
| [4] | 0,1,3 | 5 | 8 | 1x[-21] + 1x[-17] + 1x[-14] + 1x[-11] + 1x[2] + 1x[3] + 1x[5] + 1x[7] + 1x[11] + 1x[14] + 1x[17] + 1x[21] |
| [5] | 0,0,4 | 8 | 4 | 1x[-30] + 2x[-21] + 1x[-14] + 2x[4] + 2x[8] + 1x[14] + 2x[21] + 1x[30] |
| [6] | 0,3,3 | 9 | 4 | 2x[-31] + 2x[-17] + 1x[3] + 2x[7] + 2x[17] + 2x[31] *+ 1x[∞]* |
| [7] | 0,2,4 | 10 | 8 | 1x[-31] + 1x[-21] + 1x[-17] + 1x[4] + 1x[6] + 1x[8] + 1x[17] + 1x[21] + 1x[31] *+ 3x[∞]* |
| [8] | 0,1,5 | 13 | | 1x[-30] + 1x[-21] + 1x[5] + 1x[7] + 1x[21] + 1x[30] *+ 6x[∞]* |
| | | | | |
| [9] | 1,0,1 | 1 | 4 | 1x[0] + 2x[1] + 1x[2] + 2x[9] + 1x[10] + 2x[11] + 2x[12] + 1x[13] |
| [10] | 2,0,0 | 2 | 1 | 4x[9] + 4x[12] + 4x[15] |
| [11] | 1,1,2 | 3 | 8 | 1x[1] + 1x[2] + 1x[3] + 1x[4] + 1x[9] + 1x[11] + 1x[12] + 1x[13] +1x[14] + 1x[16] + 1x[17] + 1x[18] |
| [12] | 2,1,1 | 3 | 4 | 2x[9] + 1x[10] + 2x[11]+ 2x[13] + 2x[15] + 1x[16] + 2x[19] |
| [13] | 2,0,2 | 4 | 4 | 1x[9] + 2x[11] + 2x[12] + 1x[14] + 1x[15] + 2x[18] + 2x[19] + 1x[22] |
| [14] | 1,0,3 | 5 | 4 | 1x[2] + 2x[4] + 1x[5] + 2x[11] + 1x[13] + 2x[18] + 2x[21] + 1x[24] |
| [15] | 3,0,1 | 5 | 4 | 1x[10]+ 2x[12] +1x[13] + 2x[15] + 2[19] + 1x[20] + 2x[23] + 1x[25] |
| [16] | 2,2,2 | 6 | 4 | 2x[11] + 1x[12] + 2x[17] + 2x[18] + 2x[19] + 1x[26] + 2x[27] |
| [17] | 1,2,3 | 7 | 8 | 1x[3]+1x[4]+1x[6]+1x[7] + 1x[11] + 1x[16] + 1x[17] + 1x[18] + 1x[21] + 1x[26] + 1x[28] + 1x[31] |
| [18] | 2,1,3 | 7 | 8 | 1x[11] + 1x[13] + 1x[14] + 1x[16] + 1x[17] + 1x[19] + 1x[21] + 1x[22] + 1x[24] + 1x[27] + 1x[28] + 1x[32] |
| [19] | 3,1,2 | 7 | 8 | 1x[12] + 1x[13] +1x[15] + 1x[16] + 1x[18] + 1x[19] + 1x[22] + 1x[23] + 1x[25] + 1x[27] + 1x[ 29] + 1x[33] |
| [20] | 4,0,0 | 8 | 1 | 4x[15] + 4x[23] + 4x[34] |
| [21] | 1,1,4 | 9 | 8 | 1x[4]+1x[5]+1x[7] + 1x[8] + 1x[14] + 1x[17] + 1x[18] + 1x[24] + 1x[28] + 1x[30] *+2x[∞]* |
| [22] | 3,0,3 | 9 | 4 | 1x[13] + 2x[18] + 2x[19] + 1x[24] + 1x[25] + 2x[32] + 2x[33] *+ 1x[∞]* |
| [23] | 4,1,1 | 9 | 4 | 2x[15] + 2x[19] + 1x[20] + 2x[25] +1x[29] + 2x[34] *+ 2x[∞]* |
| [24] | 2,0,4 | 10 | 4 | 1x[14] + 2x[18] + 2x[21] + 1x[22] + 1x[30] + 2x[32] *+ 3x[∞]* |
| [25] | 4,0,2 | 10 | 4 | 1x[15] + 2x[19] + 1x[22] + 2x[23] + 2x[33] + 1x[34] *+ 3x[∞]* |
| [26] | 2,3,3 | 11 | 4 | 1x[16] + 2x[17] + 2x[27] + 2x[28] + 2x[31] *+ 3x[∞]* |
| [27] | 3,2,3 | 11 | 4 | 1x[16] + 1x[18] + 1x[19] + 1x[26] + 1x[27] + 1x[28] + 1x[29] + 1x[32] + 1x[33] *+ 3x[∞]* |
| [28] | 2,2,4 | 12 | 4 | 1x[17] + 1x[18] + 1x[21] + 1x[26] + 1x[27] + 1x[31] + 1x[32] *+ 5x[∞]* |
| [29] | 4,2,2 | 12 | 4 | 2x[19] + 1x[23] + 2x[27] + 2x[33] *+ 5x[∞]* |
| [30] | 1,0,5 | 13 | 4 | 1x[5] + 2x[8] + 2x[21] + 1x[24] *+ 6x[∞]* |
| [31] | 1,3,4 | 13 | 8 | 1x[6] + 1x[7] + 1x[17] + 1x[26] + 1x[28] + 1x[31] *+6x[∞]* |
| [32] | 3,1,4 | 13 | 8 | 1x[18] + 1x[22] + 1x[24] + 1x[27] + 1x[28] + 1x[33] *+ 6x[∞]* |
| [33] | 4,1,3 | 13 | 8 | 1x[19] + 1x[22] + 1x[25] + 1x[27] + 1x[29] + 1x[32] *+ 6x[∞]* |
| [34] | 5,0,1 | 13 | 4 | 1x[20] + 2x[23] + 1x[25] + 2x[34] *+ 6x[∞]* |

Table B2: The neighbouring subsets in a FCC lattice with a solute-vacancy interaction range up to the 7th neighbour shell. The distance of site $(i_1,i_2,i_3)$ to origin is expressed by the number of the neighbour shell it belongs to.

# APPENDIX C
# Alternative formulation of doubly transformed transport equation

We determine below the expression of the multiplicative coefficient $coef(m)$ of $LL_m^+$ appearing in the general equation (15). Starting from equation (14), the summation $\sum_{j=m_{sub}}^{M_{sub}} \frac{W_O f_j}{Denom} \sum_{kVj} nlink_{j\to k}\left[-LL_j^+ \, W'_{j\to k} + LL_k^+ \, W'_{k\to j}\right]$ is partitionned into two parts :

*the first part corresponds to j=m and is given by :

$$\frac{W_O f_m}{Denom} \sum_{kVm} nlink_{m\to k}\left(-LL_m^+ W'_{m\to k} + LL_k^+ W'_{k\to m}\right)$$

Due to lattice connectivity, it may happen that one possible value of ‘k’ equals ‘m’. But the two terms in the brackets cancel out each other. This is the reason why the term $LL_k^+ W'_{k\to j}$ can be dropped since non-zero contributions correspond always to cases where ‘k’ is different from ‘m’. The first part then becomes:

$$= -\frac{W_O f_m}{Denom} \sum_{kVm} nlink_{m\to k} LL_m^+ W'_{m\to k} = -\frac{W_O f_m}{Denom} \sum_{jVm} nlink_{m\to j} LL_m^+ W'_{m\to j} \, . \qquad \text{(C1)}$$

*the second part deals with the remaining available values of ‘j’ with the operator $\delta(j-m)$ and extracts all the terms corresponding to k=m with the operator $\delta(k-m)$:

$$\sum_{j=m_{sub}}^{M_{sub}} \frac{W_O f_j (1-\delta(j-m))}{Denom} \delta(k-m) \, nlink_{j\to k}\left(-LL_j^+ W'_{j\to k} + LL_k^+ W'_{k\to j}\right).$$

The only possible values of ‘j’ are dictated by a non-zero value of $nlink_{j\to m}$, which means that ‘j’ must belong to the neighbors of ‘m’. The summation $\sum_{j=m_{sub}}^{M_{sub}}$ is thus reduced to $\sum_{jVm}$. The second part then becomes:

$$\sum_{jVm} \frac{W_O f_j (1-\delta(j-m))}{Denom} nlink_{j\to m}\left(-LL_j^+ W'_{j\to m} + LL_m^+ W'_{m\to j}\right).$$

At last the operator $\delta(j-m)$ was inserted to extract the only contributions for which j≠m. Hence the term $-LL_j^+ W'_{j\to m}$ which is related to $-LL_j^+$ is not a contribution entering the coefficient of $LL_m^+$; it must be dropped too and the second part becomes:

$$\sum_{jVm} \frac{W_O f_j}{Denom} nlink_{j\to m}\left(LL_m^+ W'_{m\to j}\right). \qquad \text{(C2)}$$

The final expression of $coef(m)$ is then:

$$coef(m) = -\frac{W_O}{Denom} \sum_{jVm} W'_{m\to j}\left(f_m nlink_{m\to j} - f_j \, nlink_{j\to m}\right). \qquad \text{(C3)}$$

# Appendix D
# Proof of identities

## General expression

The basic reason for the existence of such identities is nothing but the mere fact that if site 'i' has z first neighbours 'j', each of the z neighbours 'j' has conversely site 'i' as a first neighbour. This neighbourhood relationship between the sites is transferred to a neighbourhood relationship between the subsets they belong to.
At first sight the formulation with the array $nlink_{j\to k}$ complicates the simple proof; this is the reason why we come back to the starting equation, where a single specific site $R_i$ is assumed. We first focus our attention on the corrections induced in the transport equation. These corrections consist in the following:

* first correction: the escape frequencies from site $R_i$ towards its neighbors are no longer equal to $W_0$:

$Corr1 = \delta(r-R_i)W_0\left(-\sum_{\{\omega_j\}} L(R_i,t)W'_{R_i\to R_i+\omega_j}\right)$ where $\{\omega_j\}$ is the set of first neighbour vectors;

* second correction: the entrance frequencies on the first neighbours $R_j$ coming from the specific site $R_i$ are no longer equal to $W_0$:

$$Corr2 = W_0\left(\sum_{\{\omega_j\}} \delta(r-R_i-\omega_j)L(R_i,t)W'_{R_i\to R_i+\omega_j}\right).$$

Taking the Fourier and Laplace transforms of $Corr1+Corr2$ yields a total contribution given by:

$$W_0LL(R_i,p)\sum_{\{\omega_j\}} W'_{R_i\to R_i+\omega_j}\left(-e^{-ikR_i}+e^{-ik(R_i+\omega_j)}\right) = W_0LL(R_i,p)e^{-ikR_i}\sum_{\{\omega_j\}} W'_{R_i\to R_i+\omega_j}\left(-1+e^{-ik\omega_j}\right)$$

Let us assume now that site $R_i$ belongs to a subset $\{R_i\}=\{R_i^+\}\cup\{R_i^-\}$; summing up all the contributions of sites belonging to subset 'i' gives :

$$\begin{aligned}&\sum_{\{R_i\}} e^{-ikR_i}W_0LL(R_i,p)\sum_{\{\omega_j\}} W'_{R_i\to R_i+\omega_j}(-1+e^{-ik\omega_j})\\ &= -W_0LL(R_i,p)\sum_{\{R_i\}} e^{-ikR_i}\sum_{\{\omega_j\}} W'_{R_i\to R_i+\omega_j} + W_0LL(R_i,p)\sum_{\{R_i\}}\sum_{\{\omega_j\}} W'_{R_i\to R_i+\omega_j}e^{-ik(R_i+\omega_j)}\end{aligned} \qquad \text{(D1)}$$

In the first term of Eq. D1 (right hand side), for each $R_i$ the summation over $\{\omega_j\}$ scans the z neighbours $\{R_k\}=R_i+\{\omega_j\}$ which belong necessarily to subsets 'k' referenced as neighbouring subsets $kVi$; the frequency $W'_{R_i\to R_i+\omega_j}$ depends now only on the indexes of the subsets and is renamed $W'^{sub}_{i\to k}$. The summation over $\{\omega_j\}$ can be replaced by a summation over k, taking into account the number of times $nlink_{i\to k}$ where each particular value of $W'^{sub}_{i\to k}$ is scanned; hence:

$$-W_0 LL(R_i, p) \sum_{\{R_i\}} e^{-ikR_i} \left( \sum_{\{kVi\}} nlink_{i\to k}\; W'^{sub}_{i\to k} \right)$$

Due to the underlying symmetry at the root of the definition of subsets, the same number $nlink_{i\to k}$ is invoked for each item $R_i$ of the set $\{R_i\}$; the summation over $kVi$ is the same for all items of $\{R_i\}$ and thus can be factorized. One is left with the summation of $e^{-ikR_i}$ over $\{R_i\}$ which yields the function $f_i$ (Eq. 14 of the main section). Hence the final expression of the first term:

$$-W_0 LL(R_i, p) f_i \sum_{\{kVi\}} nlink_{i\to k}\; W'^{sub}_{i\to k} \,. \qquad \text{(D2)}$$

In the second term of Eq. D1, the double summation over $\{R_i\}$ and $\{\omega_j\}$ produces vectors $\{R_k\} = \{R_i + \omega_j\}$ which belong exclusively to subsets $kVi$: it is thus possible to transform the double summation into a single summation over $\{R_k\}$. It is easy to check that a particular vector $R_k$ is generated several times when starting from a given site of subset 'i' and that the multiplicity of the enumeration is nothing but $nlink_{k\to i}$. Hence:

$$\sum_{\{R_i\}} \sum_{\{\omega_j\}} W'_{R_i \to R_i+\omega_j} e^{-ik(R_i+\omega_j)} = \sum_{\{R_k\}} e^{-ikR_k} \left( \sum_{kVi} W'^{sub}_{i\to k} nlink_{k\to i} \right)$$

Thanks to symmetries at the root of the definition of subsets, all the sites belonging to subset 'k' are equivalent and the same multiplicity $nlink_{k\to i}$ holds for any of them. Hence the summation in parenthesis is the same for all items of $\{R_k\}$ and can be factorized to yield as above

$$\left( \sum_{kVi} W'^{sub}_{i\to k} nlink_{k\to i} \right) \sum_{\{R_k\}} e^{-ikR_k} = \sum_{kVi} W'^{sub}_{i\to k} nlink_{k\to i} f_k \,. \qquad \text{(D3)}$$

Hence the expression of the total correction in the equation manipulating subset indexes after replacing the dummy index 'k' by 'j':

$$-W_0 LL(R_i, p) \sum_{jVi} W'^{sub}_{i\to j} \left( nlink_{i\to j} f_i - nlink_{j\to i} f_j \right) . \qquad \text{(D4)}$$

**A particular case of interest**

When the escape frequencies $W'^{sub}_{i\to j}$ from site $\{R_i\}$ do not depend on 'j', i.e. $W'^{sub}_{i\to j} = W'^{sub}_{i\to out}$, the contribution of sites $\{R_i\}$ belonging to subset 'i' to the doubly transformed transport equation reads :

$$-W_0 LL(R_i, p)\; W'^{sub}_{i\to out}\; (z f_i - \sum_{jVi} nlink_{j\to i} f_j) , \qquad \text{(D5)}$$

because of the relation $\sum_{jVi} nlink_{i\to j} = z$.

On the other end, the original expression can be written as well:

$$\sum_{\{R_i\}} e^{-ikR_i} W_0 LL(R_i,p) W'_{R_i\to out} \sum_{\{\omega_j\}} (-1+e^{-ik\omega_j})$$
$$= W_0 LL(R_i,p) W'_{R_i\to out} \sum_{\{R_i\}} e^{-ikR_i} \sum_{\{\omega_j\}} (-1+e^{-ik\omega_j})$$
$$= -W_0 LL(R_i,p) W'^{sub}_{i\to out} f_i D_0$$

Comparing the two expressions yields the identitiy:

$$f_i D_0 = z f_i - \sum_{jVi} nlink_{j\to i} f_j \,. \qquad \text{(D6)}$$

**Orthogonality of functions $f_i$ : evaluation of integral $-\frac{1}{n_i V_{ZB}} \int_{V_{ZB}} f_i f_m d_3 k$**

$n_i^+$ is the number of vectors in the set $\{R_i^+\}$ belonging to subset 'i' along the positive x-side and $n_i^-$ is the number of vectors $\{R_i^-\} = -\{R_i^+\}$ belonging to subset 'i' along the negative x-side; they can be grouped pairwise with opposite signs and $n_i = n_i^+ + n_i^-$.

By definition

$$f_i = \sum_{\{R_i^+\}} e^{-ikR_i^+} - \sum_{\{R_i^-\}} e^{-ikR_i^-} = \sum_{\{R_i^+\}} e^{-ikR_i^+} - \sum_{\{R_i^+\}} e^{+ikR_i^+} = -2i \sum_{\{R_i^+\}} \sin(kR_i^+)$$

Enumerating the vectors $\{R_i^+\}$ with the help of two indexes 'j' and 'k' yields :

$$f_i^2 = -4 \sum_{\{R_{ij}^+\}} \sum_{\{R_{ik}^+\}} \sin(kR_{ij}^+)\sin(kR_{ik}^+)$$
$$= -4 \sum_{\{R_{ij}^+\}} \sum_{\{R_{ik}^+\}} \left( -\frac{1}{2}\cos\left(k(R_{ij}^+ + R_{ik}^+)\right) + \frac{1}{2}\cos\left(k(R_{ij}^+ - R_{ik}^+)\right) \right)$$

The only terms bringing a non-zero contribution to the integral are those for which the argument of the cosine function is zero; this is impossible for the first term since the vectors are on the same positive x-side of the origin. For the second term, there are only $n_i^+$ contributions equal to +1 obtained for j=k. Hence

$$-\frac{1}{n_i V_{ZB}} \int_{V_{ZB}} f_i f_i d_3 k = \frac{4}{n_i V_{ZB}} \int_{V_{ZB}} \sum_{\{R_{ij}^+\}} \sum_{\{R_{ik}^+\}} \left( -\frac{1}{2}\cos\left(k(R_{ij}^+ + R_{ik}^+)\right) + \frac{1}{2}\cos\left(k(R_{ij}^+ - R_{ik}^+)\right) \right) d_3 k$$
$$= \frac{4}{n_i} \sum_{\{R_{ij}^+\}} \sum_{\{R_{ik}^+\}} \frac{1}{V_{ZB}} \int_{V_{ZB}} \left( \frac{1}{2}\cos\left(k(R_{ij}^+ - R_{ik}^+)\right) \right) d_3 k$$
$$= \frac{4}{n_i}\left(\frac{n_i^+}{2}\right) = \frac{4}{2n_i^+}\left(\frac{n_i^+}{2}\right) = 1$$

For i≠m the arguments of the cosine functions never vanish and the integral is zero.

Hence the general result $-\frac{1}{n_i V_{ZB}} \int_{V_{ZB}} f_i f_m d_3 k = \delta(i-m)$. (D7)

In Table D1-D2 below are gathered the identities and the induced relationships between lattice integrals for BCC and FCC lattices. Although the functions $f_j = f_{j_1 j_2 j_3}$ are combinations of trigonometric functions and, as such, independent from any consideration of lattice symmetry, we keep nonetheless the distinction between the two lattices for clarity since the integrals do not have the same denominator in the two lattices.

| Identities between functions $f_{j_1 j_2 j_3}$ ➔ induced relationships between lattice integrals $fifj_{i_1 i_2 i_3 \times j_1 j_2 j_3}$ ($i_1, i_2, i_3$ and $j_1, j_2, j_3$ all odd or all even) |
|---|
| $8 f_{111} = 4 f_{200} + 2 f_{202} + f_{222}$ ➔ $8\, fifj_{i_1 i_2 i_3 \times 111} = 4\, fifj_{i_1 i_2 i_3 \times 200} + 2\, fifj_{i_1 i_2 i_3 \times 202} + fifj_{i_1 i_2 i_3 \times 222} + \delta_{i_1 i_2 i_3 - 111}$ |
| $8 f_{200} = f_{111} + f_{311}$ ➔ $8\, fifj_{i_1 i_2 i_3 \times 200} = fifj_{i_1 i_2 i_3 \times 111} + fifj_{i_1 i_2 i_3 \times 311} + \delta_{i_1 i_2 i_3 - 200}$ |
| $8 f_{202} = 2 f_{111} + f_{113} + 2 f_{311} + f_{313}$ ➔ $8\, fifj_{i_1 i_2 i_3 \times 202} = 2\, fifj_{i_1 i_2 i_3 \times 111} + fifj_{i_1 i_2 i_3 \times 113} + 2\, fifj_{i_1 i_2 i_3 \times 311} + fifj_{i_1 i_2 i_3 \times 313} + \delta_{i_1 i_2 i_3 - 202}$ |
| $8 f_{113} = 2 f_{202} + 2 f_{222} + 2 f_{204} + f_{224}$ ➔ $8\, fifj_{i_1 i_2 i_3 \times 113} = 2\, fifj_{i_1 i_2 i_3 \times 202} + 2\, fifj_{i_1 i_2 i_3 \times 222} + 2\, fifj_{i_1 i_2 i_3 \times 204} + fifj_{i_1 i_2 i_3 \times 224} + \delta_{i_1 i_2 i_3 - 113}$ |
| $8 f_{311} = 4 f_{200} + 2 f_{202} + f_{222} + 4 f_{400} + 2 f_{402} + f_{422}$ ➔ $8\, fifj_{i_1 i_2 i_3 \times 311} = 4\, fifj_{i_1 i_2 i_3 \times 200} + 2\, fifj_{i_1 i_2 i_3 \times 202} + fifj_{i_1 i_2 i_3 \times 222} + 4\, fifj_{i_1 i_2 i_3 \times 400} + 2\, fifj_{i_1 i_2 i_3 \times 402} + fifj_{i_1 i_2 i_3 \times 422} + \delta_{i_1 i_2 i_3 - 311}$ |
| $8 f_{222} = f_{111} + f_{113} + f_{311} + f_{133} + f_{313} + f_{333}$ ➔ $8\, fifj_{i_1 i_2 i_3 \times 222} = fifj_{i_1 i_2 i_3 \times 111} + fifj_{i_1 i_2 i_3 \times 113} + fifj_{i_1 i_2 i_3 \times 311} + fifj_{i_1 i_2 i_3 \times 133} + fifj_{i_1 i_2 i_3 \times 313} + fifj_{i_1 i_2 i_3 \times 333} + \delta_{i_1 i_2 i_3 - 222}$ |
| $8 f_{400} = f_{311} + f_{511}$ ➔ $8\, fifj_{i_1 i_2 i_3 \times 400} = fifj_{i_1 i_2 i_3 \times 311} + fifj_{i_1 i_2 i_3 \times 511} + \delta_{i_1 i_2 i_3 - 400}$ |

Table D1: Relationships between functions and lattice integrals in a BCC lattice.

It is worth mentioning that in Table 1 of [7], the second line devoted to FCC lattice is faulty and should be replaced by:

$$11\, fifj_{101\times 112} = 2\, fifj_{101\times 101} + 2\, fifj_{101\times 211} + 2\, fifj_{101\times 202} + 2\, fifj_{101\times 103} + 2\, fifj_{101\times 222} + fifj_{101\times 123} + fifj_{101\times 213}$$

as displayed below in the third line of Table D2 devoted to the FCC lattice.

| Identities between functions $f_{j_1 j_2 j_3}$<br>➔ induced relationships between lattice integrals $fifj_{i_1 i_2 i_3 \times j_1 j_2 j_3}$<br>($i_1 + i_2 + i_3$ and $j_1 + j_2 + j_3$ even) |
| --- |
| $10\, f_{101} = 4\, f_{200} + f_{112} + 2\, f_{211} + f_{202}$<br>➔ $10\, fifj_{i_1 i_2 i_3 \times 101} = 4\, fifj_{i_1 i_2 i_3 \times 200} + fifj_{i_1 i_2 i_3 \times 112} + 2\, fifj_{i_1 i_2 i_3 \times 211} + fifj_{i_1 i_2 i_3 \times 202} + \delta_{i_1 i_2 i_3 - 101}$ |
| $12\, f_{200} = f_{101} + f_{211} + f_{301}$<br>➔ $12\, fifj_{i_1 i_2 i_3 \times 200} = fifj_{i_1 i_2 i_3 \times 101} + fifj_{i_1 i_2 i_3 \times 211} + fifj_{i_1 i_2 i_3 \times 301} + \delta_{i_1 i_2 i_3 - 200}$ |
| $11\, f_{112} = 2\, f_{101} + 2\, f_{211} + 2\, f_{202} + 2\, f_{103} + 2\, f_{222} + f_{123} + f_{213}$<br>➔ $11\, fifj_{i_1 i_2 i_3 \times 112} = 2\, fifj_{i_1 i_2 i_3 \times 101} + 2\, fifj_{i_1 i_2 i_3 \times 211} + 2\, fifj_{i_1 i_2 i_3 \times 202} + 2\, fifj_{i_1 i_2 i_3 \times 103} + 2\, fifj_{i_1 i_2 i_3 \times 222} + fifj_{i_1 i_2 i_3 \times 123} + fifj_{i_1 i_2 i_3 \times 213} + \delta_{i_1 i_2 i_3 - 112}$ |
| $12\, f_{211} = 2\, f_{101} + 4\, f_{200} + f_{112} + 2\, f_{202} + 2\, f_{301} + f_{222} + f_{312}$<br>➔ $12\, fifj_{i_1 i_2 i_3 \times 211} = 2\, fifj_{i_1 i_2 i_3 \times 101} + 4\, fifj_{i_1 i_2 i_3 \times 200} + fifj_{i_1 i_2 i_3 \times 112} + 2\, fifj_{i_1 i_2 i_3 \times 202} + 2\, fifj_{i_1 i_2 i_3 \times 301} + fifj_{i_1 i_2 i_3 \times 222} + fifj_{i_1 i_2 i_3 \times 312} + \delta_{i_1 i_2 i_3 - 211}$ |
| $12\, f_{202} = f_{101} + f_{112} + 2\, f_{211} + f_{103} + f_{301} + f_{213} + f_{312} + f_{303}$<br>➔ $12\, fifj_{i_1 i_2 i_3 \times 202} = fifj_{i_1 i_2 i_3 \times 101} + fifj_{i_1 i_2 i_3 \times 112} + 2\, fifj_{i_1 i_2 i_3 \times 211} + fifj_{i_1 i_2 i_3 \times 103} + fifj_{i_1 i_2 i_3 \times 301} + fifj_{i_1 i_2 i_3 \times 213} + fifj_{i_1 i_2 i_3 \times 312} + fifj_{i_1 i_2 i_3 \times 303} + \delta_{i_1 i_2 i_3 - 202}$ |
| $12\, f_{103} = f_{112} + f_{202} + f_{213} + f_{114} + f_{204}$<br>➔ $12\, fifj_{i_1 i_2 i_3 \times 103} = fifj_{i_1 i_2 i_3 \times 112} + fifj_{i_1 i_2 i_3 \times 202} + fifj_{i_1 i_2 i_3 \times 213} + fifj_{i_1 i_2 i_3 \times 114} + fifj_{i_1 i_2 i_3 \times 204} + \delta_{i_1 i_2 i_3 - 103}$ |
| $10\, f_{301} = 4\, f_{200} + 2\, f_{211} + f_{202} + f_{312} + 4\, f_{400} + 2\, f_{411} + f_{402}$<br>➔ $10\, fifj_{i_1 i_2 i_3 \times 301} = 4\, fifj_{i_1 i_2 i_3 \times 200} + 2\, fifj_{i_1 i_2 i_3 \times 211} + fifj_{i_1 i_2 i_3 \times 202} + fifj_{i_1 i_2 i_3 \times 312} + 4\, fifj_{i_1 i_2 i_3 \times 400} + 2\, fifj_{i_1 i_2 i_3 \times 411} + fifj_{i_1 i_2 i_3 \times 402} + \delta_{i_1 i_2 i_3 - 301}$ |
| $12\, f_{222} = f_{112} + f_{211} + f_{123} + f_{213} + f_{312} + f_{233} + f_{323}$<br>➔ $12\, fifj_{i_1 i_2 i_3 \times 222} = fifj_{i_1 i_2 i_3 \times 112} + fifj_{i_1 i_2 i_3 \times 211} + fifj_{i_1 i_2 i_3 \times 123} + fifj_{i_1 i_2 i_3 \times 213} + fifj_{i_1 i_2 i_3 \times 312} + fifj_{i_1 i_2 i_3 \times 233} + fifj_{i_1 i_2 i_3 \times 323} + \delta_{i_1 i_2 i_3 - 222}$ |
| $11\, f_{123} = f_{112} + 2\, f_{222} + f_{213} + f_{114} + 2\, f_{233} + f_{224} + f_{134}$<br>➔ $11\, fifj_{i_1 i_2 i_3 \times 123} = fifj_{i_1 i_2 i_3 \times 112} + 2\, fifj_{i_1 i_2 i_3 \times 222} + fifj_{i_1 i_2 i_3 \times 213} + fifj_{i_1 i_2 i_3 \times 114} + 2\, fifj_{i_1 i_2 i_3 \times 233} + fifj_{i_1 i_2 i_3 \times 224} + fifj_{i_1 i_2 i_3 \times 134} + \delta_{i_1 i_2 i_3 - 123}$ |
| $12\, f_{213} = f_{112} + 2\, f_{202} + 2\, f_{103} + 2\, f_{222} + f_{123} + f_{312} + f_{114} + 2\, f_{303} + 2\, f_{204} + f_{323} + f_{224} + f_{314}$<br>➔ $12\, fifj_{i_1 i_2 i_3 \times 213} = fifj_{i_1 i_2 i_3 \times 112} + 2\, fifj_{i_1 i_2 i_3 \times 202} + 2\, fifj_{i_1 i_2 i_3 \times 103} + 2\, fifj_{i_1 i_2 i_3 \times 222} + fifj_{i_1 i_2 i_3 \times 123} + fifj_{i_1 i_2 i_3 \times 312} + fifj_{i_1 i_2 i_3 \times 114} + 2\, fifj_{i_1 i_2 i_3 \times 303} + 2\, fifj_{i_1 i_2 i_3 \times 204} + fifj_{i_1 i_2 i_3 \times 323} + fifj_{i_1 i_2 i_3 \times 224} + fifj_{i_1 i_2 i_3 \times 314} + \delta_{i_1 i_2 i_3 - 213}$ |
| $11 f_{312} = 2 f_{211} + 2 f_{202} + 2 f_{301} + 2 f_{222} + f_{213} + 2 f_{303} + 2 f_{411} + 2 f_{402} + f_{323} + 2 f_{422} + f_{413}$<br>➔ $11\, fifj_{i_1 i_2 i_3 \times 312} = 2\, fifj_{i_1 i_2 i_3 \times 211} + 2\, fifj_{i_1 i_2 i_3 \times 202} + 2\, fifj_{i_1 i_2 i_3 \times 301} + 2\, fifj_{i_1 i_2 i_3 \times 222}$<br>$+ fifj_{i_1 i_2 i_3 \times 213} + 2\, fifj_{i_1 i_2 i_3 \times 303} + 2\, fifj_{i_1 i_2 i_3 \times 411} + 2\, fifj_{i_1 i_2 i_3 \times 402}$<br>$+ fifj_{i_1 i_2 i_3 \times 323} + 2\, fifj_{i_1 i_2 i_3 \times 422} + fifj_{i_1 i_2 i_3 \times 413} + \delta_{i_1 i_2 i_3 - 312}$ |
| $12 f_{400} = f_{301} + f_{411} + f_{501}$<br>➔ $12\, fifj_{i_1 i_2 i_3 \times 400} = fifj_{i_1 i_2 i_3 \times 301} + fifj_{i_1 i_2 i_3 \times 411} + fifj_{i_1 i_2 i_3 \times 501} + \delta_{i_1 i_2 i_3 - 400}$ |

Table D2: Relationships between functions and lattice integrals in a FCC lattice.

## Appendix E
## One-shot evaluation of average cosine $Q$ in the BCC lattice

The 'one-shot' approximation is rough and consists in allowing the vacancy, which dissociated previously from the OSA, to perform only one return jump to the OSA. This approximation is known to yield a returning probability always smaller than the exact one together with a correlation factor always larger than the exact one: indeed, it neglects all the trajectories of the returning vacancy which are made of more jumps.
Let us define the vectors $\Omega_{100} = a(1,0,0)$, $\Omega_{010} = a(0,1,0)$, $\Omega_{001} = a(0,0,1)$ where 'a' stands for the lattice parameter and assume Y atom is at site $\lambda_{111}$. When the half-vacancy at $r = 2\lambda_{111} = \omega_{111}$ dissociates from the OSA, the Y atom slips back to lattice site $r = 0$. The vacancy pops up into seven (unequal) parts on its seven possible neighbours.

The vacancy reaches :

- the 2[nd] neighbours of the origin at $r = \{\Omega_{100}, \Omega_{010}, \Omega_{001}\}$ with a relative probability $c_3 = W_3 / (3W_3 + 3W_3^{'} + W_3^{''})$ for each of them
- the 3[rd] neighbours of the origin at $r = \{\Omega_{100} + \Omega_{010}, \Omega_{100} + \Omega_{001}, \Omega_{010} + \Omega_{001}\}$ with a relative probability $c_3^{'} = W_3^{'} / (3W_3 + 3W_3^{'} + W_3^{''})$ for each of them
- the 5[th] neighbour of the origin at $r = \Omega_{100} + \Omega_{010} + \Omega_{001} = 2\omega_{111}$ with a relative probability $c_3^{''} = W_3^{''} / (3W_3 + 3W_3^{'} + W_3^{''})$

The above ratios of frequencies are the weights to be used in the definition of the initial condition. We then allow the vacancy to perform one jump and evaluate the probability that it comes back on a first neighbour site of the Y atom:

➔ a vacancy located at the 2[nd] neighbour site $r = \Omega_{001}$ can jump back towards:

$r = \omega_{111}$ with a relative probability $W_4 / (4W_4 + 4W_5)$ and a contribution to $Q$ equal to -1;

$r = \omega_{1\bar{1}1}$ and $r = \omega_{\bar{1}11}$ with a relative probability $W_4 / (4W_4 + 4W_5)$ for each and a contribution to $Q$ equal to -1/3;

$r = \omega_{\bar{1}\bar{1}1}$ with a relative probability $W_4 / (4W_4 + 4W_5)$ and a contribution to $Q$ equal to +1/3;

➔ a vacancy located at the 3[rd] neighbour site $r = \Omega_{100} + \Omega_{001}$ can jump back towards:

$r = \omega_{111}$ with a relative probability $W_4^{'} / (2W_4^{'} + 6W_0)$ and a contribution to $Q$ equal to -1

$r = \omega_{1\bar{1}1}$ with a relative probability $W_4^{'} / (2W_4^{'} + 6W_0)$ and a contribution to $Q$ equal to -1/3

➔ a vacancy located at $r = \Omega_{100} + \Omega_{010} + \Omega_{001} = 2\omega_{111}$ can jump back towards:

$r = \omega_{111}$ with a relative probability $W_4^{"} / (W_4^{"} + 7W_0)$ and a contribution to $Q$ equal to -1.

Summing up all the contributions gives the 'one-shot' result:

$$Q^{1s} \approx \left\{ \begin{array}{l} +3c_3 \left[ \dfrac{W_4}{4W_4 + 4W_5}(-1) + \dfrac{2W_4}{4W_4 + 4W_5}(-1/3) + \dfrac{W_4}{4W_4 + 4W_5}(+1/3) \right] \\ +3c_3^{'} \dfrac{3W_3^{'}}{3W_3 + 3W_3^{'} + W_3^{"}} \left[ \dfrac{W_4^{'}}{2W_4^{'} + 6W_0}(-1) + \dfrac{W_4^{'}}{2W_4^{'} + 6W_0}(-1/3) \right] \\ +c_3^{"} \left[ \dfrac{W_4^{"}}{W_4^{"} + 7W_0}(-1) \right] \end{array} \right\}$$

$$\approx -\frac{1}{3W_3 + 3W_3^{'} + W_3^{"}} \left[ \frac{4W_3 W_4}{4W_4 + 4W_5} + \frac{4W_3^{'} W_4^{'}}{2W_4^{'} + 6W_0} + \frac{W_3^{"} W_4^{"}}{W_4^{"} + 7W_0} \right] \qquad \text{(E1)}$$

$$f_Y^{1s} \approx 1 + Q^{1s}$$

Following Barouh's approximation [2], $W_3, W_3^{"}, W_4, W_4^{"}$ are neglected with respect to the others and $W_5 = W_0$. Then the expression of $Q^{1s}$ and $f_Y^{1s}$ reduce to :

$$Q^{1s+SOB} \approx -\frac{4W_4^{'}}{3(2W_4^{'} + 6W_0)}$$

$$f_Y^{1s+SOB} \approx \frac{W_4^{'} + 9W_0}{3W_4^{'} + 9W_0} = \frac{1}{3} + \frac{6W_0}{3W_4^{'} + 9W_0} \qquad \text{(E2)}$$

Table E1 below checks the two approximations against the exact value.
The 'one-shot' approximation is basically sound and gives the correct order of magnitude; it is close to the exact result at the lower temperatures because of the overwhelming contribution of $W_4^{"}$ and departs progressively when the temperature is raised. The approximation conducted in the spirit of Barouh [2] gives a nearly temperature independent value close to 1/3 because $W_4^{'} >> W_0$ .

| T(K) | $f_Y^{exact}$ | $f_Y^{1s}$ | $f_Y^{1s+SOB}$ |
|---|---|---|---|
| 300 | 4.356 $10^{-4}$ | 4.356 $10^{-4}$ | 0.333 |
| 400 | 2.993 $10^{-3}$ | 2.993 $10^{-3}$ | 0.333 |
| 500 | 9.368 $10^{-3}$ | 9.371 $10^{-3}$ | 0.333 |
| 600 | 1.966 $10^{-2}$ | 1.969 $10^{-2}$ | 0.333 |
| 700 | 3.275 $10^{-2}$ | 3.293 $10^{-2}$ | 0.333 |
| 800 | 4.720 $10^{-2}$ | 4.787 $10^{-2}$ | 0.333 |
| 900 | 6.190 $10^{-2}$ | 6.369 $10^{-2}$ | 0.334 |
| 1000 | 7.613 $10^{-2}$ | 8.003 $10^{-2}$ | 0.334 |
| 1100 | 8.958 $10^{-2}$ | 9.686 $10^{-2}$ | 0.335 |

Table E1: Comparison between the exact value of the correlation factor for Y in BCC Fe, the 'one-shot' approximation and the approximation in the spirit of Barouh.

## Appendix F
## Correlation factor for self-diffusion of OSA with the new vacancy mechanism in BCC and FCC lattices

If all the frequencies are equal to $W_0$ the system of unknows of Eq. (40) is highly simplified; only the coefficients of the first column and of the main diagonal remain.

### BCC lattice

$$fifj(5,5)\frac{pLL_5^+}{W_0} = \frac{fifj(5,6)+fifj(5,7)/2+fifj(5,10)/4}{7W_0}$$
$$fifj(6,5)\frac{pLL_5^+}{W_0} + LL_6^+ = \frac{fifj(6,6)+fifj(6,7)/2+fifj(6,10)/4}{7W_0}$$
$$fifj(7,5)\frac{pLL_5^+}{W_0} + LL_7^+ = \frac{fifj(7,6)+fifj(7,7)/2+fifj(7,10)/4}{7W_0}$$
$$fifj(10,5)\frac{pLL_5^+}{W_0} + LL_{10}^+ = \frac{fifj(10,6)+fifj(10,7)/2+fifj(10,10)/4}{7W_0}$$

The solution is immediate:

$$\frac{pLL_5^+}{W_0} = \frac{fifj(5,6)+fifj(5,7)/2+fifj(5,10)/4}{7\,fifj(5,5)W_0}$$

and

$$LL_6^+ = \frac{fifj(6,6)+fifj(6,7)/2+fifj(6,10)/4}{7W_0} - \frac{fifj(6,5)}{fifj(5,5)}\frac{fifj(5,6)+fifj(5,7)/2+fifj(5,10)/4}{7W_0}$$
$$LL_7^+ = \frac{fifj(7,6)+fifj(7,7)/2+fifj(7,10)/4}{7W_0} - \frac{fifj(7,5)}{fifj(5,5)}\frac{fifj(5,6)+fifj(5,7)/2+fifj(5,10)/4}{7W_0}$$
$$LL_{10}^+ = \frac{fifj(10,6)+fifj(10,7)/2+fifj(10,10)/4}{7W_0} - \frac{fifj(10,5)}{fifj(5,5)}\frac{fifj(5,6)+fifj(5,7)/2+fifj(5,10)/4}{7W_0}$$

The average cosine Q is equal to:

$$Q_0^{BCC} = -4LL_6^+\,W_0 \quad -8LL_7^+\,W_0 \quad -4LL_{10}^+\,W_0$$
$$= -\frac{\left[4\,fifj(6,6)+2\,fifj(6,7)+fifj(6,10)\right]}{7} + \frac{fifj(6,5)}{fifj(5,5)}\frac{\left[4\,fifj(5,6)+2\,fifj(5,7)+fifj(5,10)\right]}{7}$$
$$-\frac{\left[8\,fifj(7,6)+4\,fifj(7,7)+2\,fifj(7,10)\right]}{7} + \frac{2\,fifj(7,5)}{fifj(5,5)}\frac{\left[4\,fifj(5,6)+2\,fifj(5,7)+fifj(5,10)\right]}{7}$$
$$-\frac{4\,fifj(10,6)+2\,fifj(10,7)+fifj(10,10)}{7} + \frac{fifj(10,5)}{fifj(5,5)}\frac{\left[4\,fifj(5,6)+2\,fifj(5,7)+fifj(5,10)\right]}{7}$$

The expression $4\,fifj(5,6)+2\,fifj(5,7)+fifj(5,10)$ can be reduced thanks to the relationships between lattice integrals. Using Table B1 for the correspondence between the subset number and the coordinates of the representative site, we get: $4\,fifj(5,6)+2\,fifj(5,7)+fifj(5,10) \equiv 4\,fifj_{111\times200}+2\,fifj_{111\times202}+fifj_{111\times222}$. From the first line of Table D1, we get :

$4\,fifj_{111\times200}+2\,fifj_{111\times202}+\,fifj_{111\times222}=8\,fifj_{111\times111}-1\equiv8\,fifj(5,5)-1$. Hence the second bracket in the three lines expressing $Q_0^{BCC}$ is equal to $8\,fifj(5,5)-1$. We get:

$$Q_0^{BCC}=-\frac{[4\,fifj(6,6)+2\,fifj(6,7)+\,fifj(6,10)-8\,fifj(6,5)]}{7}-\frac{fifj(6,5)}{7\,fifj(5,5)}$$
$$-\frac{[8\,fifj(7,6)+4\,fifj(7,7)+2\,fifj(7,10)-16\,fifj(7,5)]}{7}-\frac{2\,fifj(7,5)}{7\,fifj(5,5)}$$
$$-\frac{[4\,fifj(10,6)+2\,fifj(10,7)+\,fifj(10,10)-8\,fifj(10,5)]}{7}-\frac{fifj(10,5)}{7\,fifj(5,5)}$$

The same Table D1 can be used to show that the three remaining square brackets are identically zero. Hence:

$$Q_0^{BCC}=-\frac{fifj(6,5)+2\,fifj(7,5)+\,fifj(10,5)}{7\,fifj(5,5)}=-0.2383969. \qquad \text{(F1)}$$

The integrals are calculated with a superposition of three Gaussian quadratures along the three axes as indicated in [7]. The correlation factor for self-diffusion with this mechanism is found equal to

$$f_0=1+Q_0^{BCC}=0.7616031. \qquad \text{(F2)}$$

The fact that correlation effects are strong in the case of Y in Fe is thus not intrinsically related to the mechanism itself but only to the strong attractive interactions between the vacancy and the Y atom and to the resulting pattern of vacancy jump frequencies around the Y, namely small frequencies for dissociating the complex (V/2+OSA+V/2) and large frequencies to build it.

## FCC lattice

$$fifj(9,9)\frac{pLL_9^+}{W_0}=\frac{4\,fifj(9,10)+\,fifj(9,11)+2\,fifj(9,12)+\,fifj(9,13)}{28W_0}$$
$$fifj(10,9)\frac{pLL_9^+}{W_0}+LL_{10}^+=\frac{4\,fifj(10,10)+\,fifj(10,11)+2\,fifj(10,12)+\,fifj(10,13)}{28W_0}$$
$$fifj(11,9)\frac{pLL_9^+}{W_0}+LL_{11}^+=\frac{4\,fifj(11,10)+\,fifj(11,11)+2\,fifj(11,12)+\,fifj(11,13)}{28W_0}$$
$$fifj(12,9)\frac{pLL_9^+}{W_0}+LL_{12}^+=\frac{4\,fifj(12,10)+\,fifj(12,11)+2\,fifj(12,12)+\,fifj(12,13)}{28W_0}$$
$$fifj(13,9)\frac{pLL_9^+}{W_0}+LL_{13}^+=\frac{4\,fifj(13,10)+\,fifj(13,11)+2\,fifj(13,12)+\,fifj(13,13)}{28W_0}$$

Hence:

$$\frac{pLL_9^+}{W_0}=\frac{4\,fifj(9,10)+\,fifj(9,11)+2\,fifj(9,12)+\,fifj(9,13)}{28\,fifj(9,9)W_0}$$

which implies:

$$LL_{10}^{+} = \frac{4\,fifj(10,10) + fifj(10,11) + 2\,fifj(10,12) + fifj(10,13)}{28W_0} - fifj(10,9)\frac{4\,fifj(9,10) + fifj(9,11) + 2\,fifj(9,12) + fifj(9,13)}{28\,fifj(9,9)W_0}$$

$$LL_{11}^{+} = \frac{4\,fifj(11,10) + fifj(11,11) + 2\,fifj(11,12) + fifj(11,13)}{28W_0} - fifj(11,9)\frac{4\,fifj(9,10) + fifj(9,11) + 2\,fifj(9,12) + fifj(9,13)}{28\,fifj(9,9)W_0}$$

$$LL_{12}^{+} = \frac{4\,fifj(12,10) + fifj(12,11) + 2\,fifj(12,12) + fifj(12,13)}{28W_0} - fifj(12,9)\frac{4\,fifj(9,10) + fifj(9,11) + 2\,fifj(9,12) + fifj(9,13)}{28\,fifj(9,9)W_0}$$

$$LL_{13}^{+} = \frac{4\,fifj(13,10) + fifj(13,11) + 2\,fifj(13,12) + fifj(13,13)}{28W_0} - fifj(13,9)\frac{4\,fifj(9,10) + fifj(9,11) + 2\,fifj(9,12) + fifj(9,13)}{28\,fifj(9,9)W_0}$$

Thanks to the relationships between integrals which can be extracted from Table D2, we obtain as above:

$$4\,fifj(9,10) + fifj(9,11) + 2\,fifj(9,12) + fifj(9,13) = 10\,fifj(9,9) - 1\,,$$
$$4\,fifj(10,10) + fifj(10,11) + 2\,fifj(10,12) + fifj(10,13) - 10\,fifj(10,9) = 0\,,$$
$$4\,fifj(11,10) + fifj(11,11) + 2\,fifj(11,12) + fifj(11,13) - 10\,fifj(11,9) = 0\,,$$
$$4\,fifj(12,10) + fifj(12,11) + 2\,fifj(12,12) + fifj(12,13) - 10\,fifj(12,9) = 0\,,$$
$$4\,fifj(13,10) + fifj(13,11) + 2\,fifj(13,12) + fifj(13,13) - 10\,fifj(13,9) = 0\,,$$

and the unknowns reduce to:

$$LL_{10}^{+} = \frac{fifj(10,9)}{28\,fifj(9,9)W_0} \qquad LL_{11}^{+} = \frac{fifj(11,9)}{28\,fifj(9,9)W_0}$$
$$LL_{12}^{+} = \frac{fifj(12,9)}{28\,fifj(9,9)W_0} \qquad LL_{13}^{+} = \frac{fifj(13,9)}{28\,fifj(9,9)W_0}.$$

The average cosine is given by:

$$\begin{aligned} Q_0^{FCC} &= -4LL_{10}^{+}\,W_0 - 8LL_{11}^{+}\,W_0 - 8LL_{12}^{+}\,W_0 - 4LL_{13}^{+}\,W_0 \\ &= -\frac{fifj(10,9) + 2\,fifj(11,9) + 2\,fifj(12,9) + fifj(13,9)}{7\,fifj(9,9)}. \\ &= -0.2737533306 \end{aligned} \qquad \text{(F3)}$$

For self-diffusion $2W_{IS} = 14W_0$ and $W_{II} = W_0$. Hence $2\alpha = 7/11$ and the correlation factor with this mechanism is equal to

$$f_0 = 1 + \frac{7}{9}Q_0^{FCC} = 0.78708074. \qquad \text{(F4)}$$

# Appendix G
# Recurrence equations for the FCC structure

The moments for the unmodified recurrence equation were already calculated in Appendix A:

$$SI_n^{\lambda_i}(r) = \sum_{\{\lambda_j\}} IS_{n-1}^{-\lambda_i}(r+\lambda_j)\, p_{-\lambda_j,\lambda_i}$$

$$\rightarrow \quad M_{SI0} - \frac{1}{z} = PM_{IS0}$$

$$M_{SI1} - \frac{1}{z} = M_{IS1}Q + PM_{IS0}$$

$$M_{SI2} - \frac{1}{z} = M_{IS2}P + 2M_{IS1}Q + M_{IS0}P$$

Moments for the new equation suited for the FCC lattice:

$$IS_n^{-\lambda_i}(r+\lambda_i) = \alpha\ SI_{n-1}^{\lambda_i}(r) + \alpha\ SI_{n-1}^{-\lambda_i}(r+2\lambda_i) + \beta \sum_{\{\lambda_j^+\}} IS_{n-1}^{-\lambda_j^+}(r+\lambda_j^+) + \beta \sum_{\{\lambda_j^-\}} IS_{n-1}^{-\lambda_j^-}(r+2\lambda_i+\lambda_j^-)$$

Because the first jump of the encounter is of S→I type, the initial condition for the IS functions is identically zero and $IS_1^{-\lambda_i}(r) = 0 \quad \forall r, \forall \lambda_i$. For the IS functions, the summation over 'n' from 2 to ∞ can thus be extended to a summation from n=1 to ∞.

*Zeroth order moments:*

$$M_{IS0} = 2\alpha M_{SI0} + 8\beta M_{IS0} \rightarrow M_{IS0}(1-8\beta) = 2\alpha M_{SI0}$$

Hence : $M_{IS0} = M_{SI0}$

*First order moments:*

LHS :

$$\sum_{\{r\}} \sum_{n=2}^{\infty} rIS_n^{-\lambda_i}(r+\lambda_i) = \sum_{\{r\}} \sum_{n=1}^{\infty} rIS_n^{-\lambda_i}(r+\lambda_i) = -M_{IS1}\overrightarrow{\lambda_i}$$

RHS : terms proportional to $\alpha$

$$\alpha \sum_{\{r\}} \sum_{n=2}^{\infty} \left((r+\lambda_i)SI_{n-1}^{\lambda_i}(r) - \lambda_i SI_{n-1}^{\lambda_i}(r)\right) = \alpha \sum_{\{r\}} \sum_{n=2}^{\infty} (r+\lambda_i)SI_{n-1}^{\lambda_i}(r) - \alpha \sum_{\{r\}} \sum_{n=2}^{\infty} \lambda_i SI_{n-1}^{\lambda_i}(r)$$

$$= \alpha(M_{SI1} - M_{SI0})\overrightarrow{\lambda_i}$$

$$\alpha \sum_{\{r\}} \sum_{n=2}^{\infty} \left((r+\lambda_i)SI_{n-1}^{-\lambda_i}(r+2\lambda_i) - \lambda_i SI_{n-1}^{-\lambda_i}(r+2\lambda_i)\right) = \alpha M_{SI1}(\overrightarrow{-\lambda_i}) - \alpha M_{SI0}(\overrightarrow{\lambda_i})$$

$$= -\alpha(M_{SI1} + M_{SI0})\overrightarrow{\lambda_i}$$

RHS : first term proportional to $\beta$

$$\beta \sum_{\{r\}} \sum_{n=2}^{\infty} \sum_{\{\lambda_j^+\}} rIS_{n-1}^{-\lambda_j^+}(r+\lambda_j^+) = -\beta \sum_{\{\lambda_j^+\}} M_{IS1}\overrightarrow{\lambda_j^+}$$

RHS : second term proportional to $\beta$

$$\beta\sum_{\{r\}}\sum_{n=2}^{\infty}\sum_{\{\lambda_j^-\}} rIS_{n-1}^{-\lambda_j^-}(r+2\lambda_i+\lambda_j^-)$$

$$=\beta\sum_{\{r\}}\sum_{n=2}^{\infty}\sum_{\{\lambda_j^-\}}(r+2\lambda_i)IS_{n-1}^{-\lambda_j^-}(r+2\lambda_i+\lambda_j^-)-2\overrightarrow{\lambda_i}\,\beta\sum_{\{r\}}\sum_{n=2}^{\infty}\sum_{\{\lambda_j^-\}}IS_{n-1}^{-\lambda_j^-}(r+2\lambda_i+\lambda_j^-)$$

$$=-\beta\sum_{\{\lambda_j^-\}}M_{IS1}\overrightarrow{\lambda_j^-}-2\overrightarrow{\lambda_i}\beta\sum_{\{\lambda_j^-\}}M_{IS0}=-\beta\sum_{\{\lambda_j^-\}}M_{IS1}\overrightarrow{\lambda_j^-}-8\beta M_{IS0}\overrightarrow{\lambda_i}$$

$$=\beta\sum_{\{\lambda_j^+\}}M_{IS1}\overrightarrow{\lambda_j^+}-8\beta M_{IS0}\overrightarrow{\lambda_i}$$

Hence the equation for the first moment :

$$-M_{IS1}\overrightarrow{\lambda_i}$$

$$=\alpha(M_{SI1}-M_{SI0})\overrightarrow{\lambda_i}-\alpha(M_{SI1}+M_{SI0})\overrightarrow{\lambda_i}-\beta\sum_{\{\lambda_j^+\}}M_{IS1}\overrightarrow{\lambda_j^+}+\beta\sum_{\{\lambda_j^+\}}M_{IS1}\overrightarrow{\lambda_j^+}-8\beta M_{IS0}\overrightarrow{\lambda_i}$$

$$=-(2\alpha+8\beta)M_{SI0}\overrightarrow{\lambda_i}=-M_{SI0}\overrightarrow{\lambda_i}$$

Hence $M_{IS1}=M_{SI0}$

*Second order moments :*

LHS:

$$\sum_{\{r\}}\sum_{n=2}^{\infty}r^2IS_n^{-\lambda_i}(r+\lambda_i)=\sum_{\{r\}}\sum_{n=1}^{\infty}r^2IS_n^{-\lambda_i}(r+\lambda_i)=M_{IS2}\lambda^2$$

RHS : first term proportional to $\alpha$

$$\alpha\sum_{\{r\}}\sum_{n=2}^{\infty}r^2SI_{n-1}^{\lambda_i}(r)$$

$$=\alpha\sum_{\{r\}}\sum_{n=2}^{\infty}\left[(r+\lambda_i)^2-2\lambda_i(r+\lambda_i)+\lambda_i^2\right]SI_{n-1}^{\lambda_i}(r)$$

$$=\alpha\sum_{\{r\}}\sum_{n=2}^{\infty}\left[(r+\lambda_i)^2\right]SI_{n-1}^{\lambda_i}(r)-\alpha\sum_{\{r\}}\sum_{n=2}^{\infty}\left[2\lambda_i(r+\lambda_i)\right]SI_{n-1}^{\lambda_i}(r)+\alpha\sum_{\{r\}}\sum_{n=2}^{\infty}\left[\lambda_i^2\right]SI_{n-1}^{\lambda_i}(r)$$

$$=\alpha M_{SI2}\lambda^2-2\alpha(M_{SI1}\overrightarrow{\lambda_i}).\overrightarrow{\lambda_i}+\alpha M_{SI0}\lambda^2$$

$$=\alpha(M_{SI2}-2M_{SI1}+M_{SI0})\lambda^2$$

RHS : second term proportional to $\alpha$

$$\alpha\sum_{\{r\}}\sum_{n=2}^{\infty}r^2SI_{n-1}^{-\lambda_i}(r+2\lambda_i)=\alpha\sum_{\{r\}}\sum_{n=2}^{\infty}\left[(r+\lambda_i)^2-2\lambda_i(r+\lambda_i)+\lambda_i^2\right]SI_{n-1}^{-\lambda_i}(r+2\lambda_i)$$

The same calculation as above, while changing the sign of the first moment, yields :

$$\alpha\sum_{\{r\}}\sum_{n=2}^{\infty}r^2SI_{n-1}^{-\lambda_i}(r+2\lambda_i)=\alpha(M_{SI2}+2M_{SI1}+M_{SI0})\lambda^2$$

RHS : first term proportional to $\beta$

$$\beta\sum_{\{r\}}\sum_{n=2}^{\infty} r^2 \sum_{\{\lambda_j^+\}} IS_{n-1}^{-\lambda_j^+}(r+\lambda_j^+) = \beta\sum_{\{\lambda_j^+\}} M_{IS2}(\lambda_j^+)^2 = 4\beta M_{IS2}\lambda^2$$

RHS : second term proportional to $\beta$

$$\beta\sum_{\{\lambda_j^-\}}\sum_{\{r\}}\sum_{n=2}^{\infty} r^2 IS_{n-1}^{-\lambda_j^-}(r+2\lambda_i+\lambda_j^-)$$

$$=\beta\sum_{\{\lambda_j^-\}}\sum_{\{r\}}\sum_{n=2}^{\infty}\left[(r+2\lambda_i)^2-4\lambda_i(r+2\lambda_i)+4\lambda^2\right]IS_{n-1}^{-\lambda_j^-}(r+2\lambda_i+\lambda_j^-)$$

The first part of the square bracket :

$$=\beta\sum_{\{\lambda_j^-\}}\sum_{\{r\}}\sum_{n=2}^{\infty}\left[(r+2\lambda_i)^2\right]IS_{n-1}^{-\lambda_j^-}(r+2\lambda_i+\lambda_j^-) = 4\beta M_{IS2}\lambda^2$$

The second part of the square bracket :

$$-\beta\sum_{\{\lambda_j^-\}}\sum_{\{r\}}\sum_{n=2}^{\infty}\left[4\lambda_i(r+2\lambda_i)\right]IS_{n-1}^{-\lambda_j^-}(r+2\lambda_i+\lambda_j^-)$$

$$=-4\lambda_i\beta\sum_{\{\lambda_j^-\}}(-M_{IS1}\overrightarrow{\lambda_j^-}) = -4\lambda_i\beta\sum_{\{\lambda_j^+\}}(M_{IS1}\overrightarrow{\lambda_j^+}) = -8\beta M_{IS1}\lambda^2$$

The third part of the square bracket :

$$+\beta\sum_{\{\lambda_j^-\}}\sum_{\{r\}}\sum_{n=2}^{\infty}\left[4\lambda^2\right]IS_{n-1}^{-\lambda_j^-}(r+2\lambda_i+\lambda_j^-) = +4\beta\lambda^2(4M_{IS0}) = +16\beta M_{IS0}\lambda^2$$

which gives a total : $\rightarrow +4\beta M_{IS2}\lambda^2 - 8\beta M_{IS1}\lambda^2 + 16\beta M_{IS0}\lambda^2$

Hence the final equation:

$$\begin{aligned} M_{IS2}\lambda^2 &= \alpha(M_{SI2}-2M_{SI1}+M_{SI0})\lambda^2 + \alpha(M_{SI2}+2M_{SI1}+M_{SI0})\lambda^2 \\ &\quad + 4\beta M_{IS2}\lambda^2 + 4\beta M_{IS2}\lambda^2 - 8\beta M_{IS1}\lambda^2 + 16\beta M_{IS0}\lambda^2 \\ &= 2\alpha(M_{SI2}+M_{SI0})\lambda^2 + 8\beta M_{IS2}\lambda^2 - 8\beta M_{IS1}\lambda^2 + 16\beta M_{IS0}\lambda^2 \end{aligned}$$

or

$$2\alpha M_{IS2} = 2\alpha M_{SI2} + 2\alpha M_{SI0} - 8\beta M_{IS1} + 16\beta M_{IS0}$$

Solving the system for $M_{IS2}$ gives :

$$M_{IS1} = M_{SI0} = M_{IS0}, \tag{G1}$$

and

$$\frac{M_{IS2}}{M_{IS0}} = \frac{1+2\alpha+4\alpha Q}{2\alpha(1-P)}. \tag{G2}$$

The mean square displacement during an encounter is thus:

$$< R^2 >_{Enc} = \frac{12M_{IS2}\,\lambda^2}{12M_{IS0}} = \frac{1+2\alpha+4\alpha Q}{2\alpha(1-P)}\lambda^2 = \frac{1+2\alpha+4\alpha Q}{8\alpha(1-P)}\omega^2. \tag{G3}$$

The average quadratic length of the macrojump is obtained for $P=Q=0$:

$$< R^2 >_{MJ} = \frac{1+2\alpha}{2\alpha}\lambda^2. \tag{G4}$$

The mean random displacement is

$$< R^2 >_{Rand} = \frac{1+2\alpha}{2\alpha(1-P)}\lambda^2 , \quad \text{(G5)}$$

and the correlation factor in the FCC lattice:

$$f_B = 1 + \frac{4\alpha Q}{1+2\alpha} . \quad \text{(G6)}$$